\documentclass[onecolumn,12pt,draftcls]{IEEEtran}
\usepackage{amsmath}
\usepackage{amsfonts}
\usepackage{bbding}
\usepackage{amssymb}
\usepackage{array}
\usepackage{subfigure}

\usepackage{graphicx}
\usepackage{subfigure}
\usepackage[named]{algo}
\usepackage{algorithmic}
\usepackage{psfrag}
\usepackage{stfloats}
\usepackage[compress]{cite}
\makeatletter
\renewcommand{\citepunct}{,\penalty\@m\hskip.13emplus.1emminus.1em}
\renewcommand{\citedash}{\hbox{--}\penalty\@m}
\makeatother
\usepackage{setspace}
\usepackage{color}
\allowdisplaybreaks

\usepackage{amsthm}
\usepackage{stfloats}

\newtheorem{rem}{Remark}
\newtheorem{pro}{Property}
\newtheorem{prop}{Proposition}

\usepackage{hyperref}
\usepackage{bm}
\renewcommand{\IEEEQED}{\IEEEQEDopen}

\begin{document}
\title{Joint Uplink and Downlink Resource Configuration for Ultra-reliable and Low-latency Communications}

\author{
\IEEEauthorblockN{{Changyang She, Chenyang Yang and Tony Q.S. Quek}}

\thanks{This paper was presented in part at the workshop on ultra-reliable and low-latency communications in wireless networks with
IEEE Global Communications Conference 2016 \cite{She2016GCworkshop}.}

\thanks{Changyang She was with the School of Electronics and
	Information Engineering, Beihang University, Beijing 100191, China. He is now with the Information Systems Technology and Design Pillar, Singapore University of Technology and Design, 8 Somapah Road, Singapore 487372 (email:shechangyang@gmail.com).}

\thanks{Chenyang Yang is with the School of Electronics and
Information Engineering, Beihang University, Beijing 100191, China (email:cyyang@buaa.edu.cn).}

\thanks{T. Q. S. Quek is with the Information Systems Technology and Design
	Pillar, Singapore University of Technology and Design, 8 Somapah Road, Singapore 487372 (e-mail: tonyquek@sutd.edu.sg).} \vspace{0.0cm}}

\maketitle
\begin{abstract} Supporting ultra-reliable and low-latency communications (URLLC) is one of the major goals for the fifth-generation cellular networks. Since spectrum usage efficiency is always a concern, and large bandwidth is required for ensuring stringent quality-of-service (QoS), we minimize the total bandwidth under the QoS constraints of URLLC. We first propose a packet delivery mechanism for URLLC. To reduce the required bandwidth for ensuring queueing delay, we consider a statistical multiplexing queueing mode, where the packets to be sent to different devices are waiting in one queue at the base station, and broadcast mode is adopted in downlink transmission. In this way, downlink bandwidth is shared among packets of multiple devices. In uplink transmission, different subchannels are allocated to different devices to avoid strong interference. Then, we jointly optimize uplink and downlink bandwidth configuration and delay components to minimize the total bandwidth required to guarantee the overall packet loss and end-to-end delay, which includes uplink and downlink transmission delays, queueing delay and backhaul delay. We propose a two-step method to find the optimal solution. Simulation and numerical results validate our analysis and show remarkable performance gain by jointly optimizing uplink and downlink configuration.
\end{abstract}


\section{Introduction}
Ultra-reliable and low-latency communications (URLLC) are required in the emerging application scenarios of the fifth-generation (5G) cellular networks \cite{3GPP2016Scenarios}. Different from existing communication systems that are designed for human-to-human (H2H) communications, URLLC target to ultra-reliable machine-type-communications and human-to-machine communications that require haptic interactions, such as autonomous vehicles, factory automation, and remote control \cite{Popovski2014METIS,Gerhard2014The}. All those applications have strict requirements on end-to-end (E2E) or round trip delay (say around $1$~ms) and reliability (say around $10^{-6}$ packet loss probability), which can not be satisfied in Long Term Evolution systems.

The E2E delay consists of various delay components that depend on communication scenarios. In long distance communication scenarios, the E2E delay consists of transmission delay and queueing delay in radio access network, routing delay in backhaul and core networks, and also propagation delay that is hard to control. For example, when the communication distance is longer than $300$~km, the propagation delay is longer than $1$~ms since light travels $300$~km per millisecond in vacuum \cite{Meryem2016Tactile}. Nevertheless, it is worth noting that ensuring the ultra-short E2E delay  is not easy even in local communication scenarios, where communication is only required by users in adjacent cells with short backhaul delay and negligible propagation delay.

A core difference between radio resource allocation for URLLC and that for traditional real-time service comes from both transmission delay and packet size \cite{3GPP2016Scenarios}. In H2H communications, transmission delay is relative long (say 10 ms) and the packet size is large (say 1500 bytes) \cite{3GPPQoS}. As a result, Shannon's Capacity is widely applied in existing literatures to characterize achievable rate of traditional services with long packets (e.g., \cite{She2016TWC} and references therein). In URLLC, to satisfy short transmission delay, short packets are transmitted (say $20$~byte \cite{3GPP2016Scenarios}). As a result, the blocklength of channel coding is short. In addition, to ensure ultra-high reliability, decoding error with short blocklength channel codes cannot be ignored. Therefore, decoding error probability with short blocklength channel codes should be applied \cite{Giuseppe2016Toward}, which is with very complicated expression. Fortunately, approximate decoding error probability in finite blocklength regime has been obtained with simple expression in \cite{Yury2010Channel,Yury2014Quasi}, which are shown accurate for quasi-static fading channels. Yet these approximations are neither convex nor concave in transmit power or bandwidth. As a result, the resource allocation optimization for URLLC is much more challenging than H2H communications.

Similar to H2H communications, ensuring short queueing delay is also necessary for URLLC, where queueing delay requirement should be characterized by the queueing delay bound and its violation probability. Considering that packets are randomly generated and the service rate of a wireless link could be random, queueing delay has been considered in single-user scenarios \cite{Shengfeng2016Convexity,Gross2015Delay,Yulin2016Blocklength} and multi-user scenarios  \cite{She2017CrossLayer}, where achievable rate in finite blocklength regime was applied in their analyses. A packet scheduling policy was proposed under strict delay bound constraint on queueing delay in \cite{Shengfeng2016Convexity}, which cannot be satisfied with probability one due to channel fading. To show when the delay bound can be satisfied, a feasible condition was obtained, but delay bound violation probability can not be derived under the framework in \cite{Shengfeng2016Convexity}. To analyze queueing delay bound violation probability, network calculus was applied to obtain an upper bound of the delay violation probability in \cite{Gross2015Delay}. Simulation results in \cite{Gross2015Delay} validated that if Shannon's Capacity is applied, then the delay violation probability will be underestimated, and hence the quality-of-service (QoS) cannot be guaranteed. The performance of relay systems was analyzed in \cite{Yulin2016Blocklength}, where effective capacity was applied to characterize the queueing delay. More recently, the transmission policy in both single-user and multi-user scenarios was optimized to minimize the maximal transmit power required to satisfy QoS requirement in \cite{She2017CrossLayer}. To study how to serve multiple users with multiple BSs, a signal-to-interference-plus-noise ratio (SINR) model was applied to simplify the reliability requirement in \cite{Jie2017Availability}, where multi-connectivity was exploited to improve reliability.

The study in \cite{She2017CrossLayer} focuses on downlink (DL) transmission design, and implicitly assume that the uplink (UL) transmission can be finished in a short time with guaranteed reliability. However, ensuring ultra-reliable and low-latency for UL transmission is not easy as well. To ensure the QoS for URLLC, UL and DL resource allocation should be jointly optimized \cite{Adnan2016Towards}. To guarantee queueing delay violation probability, effective bandwidth and effective capacity were applied in \cite{Adnan2016Towards}, where the Shannon's Capacity was used as the service rate (and hence decoding error probability was not considered), and the global optimal solutions of the problem was not found. If the achievable rate in finite blocklength regime is applied in the joint UL and DL resource allocation, it could be more challenging to find the optimal solution \cite{Changyang2017Mag}.

On the other hand, spectrum is scarce resource for wireless communications. Due to stringent QoS requirement, the resource allocation for URLLC is inevitably conservative, and hence the bandwidth usage efficiency for URLLC will be very low without judicious control. In some application scenarios, the overall bandwidth can become unaffordable. For example, the packet arrival rate in tactile internet is very high \cite{Condoluci2017Soft}, and in smart factory the number of devices can be very large \cite{Philipp2017IoT}. In order to improve the spectrum usage efficiency, the packet delivery mechanism for URLLC, including queueing and transmission modes, should be reconsidered. To guarantee the QoS of each user, the packets to different users are waiting in different queues (i.e., individual queueing mode) in \cite{She2017CrossLayer,Adnan2016Towards}. However, to achieve the same average queueing delay, the required service rate of the individual queueing mode is much higher than that of a statistical multiplexing queueing mode, where the packets to different users are waiting in one common queue, and hence the DL bandwidth can be shared among multiple users \cite{Mor2013Queue}. Besides, most of existing studies assumed that channel state information at transmitter (CSIT) is perfectly known, which incurs training/feedback overhead that linearly increases with the number of receivers \cite{Shengfeng2016Convexity,Gross2015Delay,She2017CrossLayer,Adnan2016Towards}. Since the packet size in URLLC is usually very small, large signaling overhead leads to low spectrum efficiency. With a transmission mechanism without CSIT, the overhead can be reduced, but whether the QoS requirement can be guaranteed becomes a question. 

In this paper, we study how to jointly optimize UL and DL resource configuration and delay components to minimize the total bandwidth to support URLLC in  local communication scenarios. We focus on orthogonal multiple access systems to avoid strong and random interference. The major contributions are summarized as follows:
\begin{itemize}
\item We propose a packet delivery mechanism for URLLC. To save bandwidth in UL transmission, bandwidth is not reserved for all the UL devices since they may stay dumb for a long time. Only the devices that request sending packets will be assigned bandwidth. To reduce the bandwidth for ensuring queueing delay requirement, statistical multiplex queueing mode is adopted, where the packets to different users are waiting in one queue at the buffer of the base station (BS). By taking Poisson arrival process as an example, we prove that under the same queueing delay bound and queueing delay violation probability, the required service rate of a statistical multiplex queue is less than the sum of the required service rates of all individual queues, where the packets to different users are waiting in different queues. To reduce overhead, broadcast is applied for DL transmission.
\item We jointly optimize the bandwidth assignment for UL and DL transmissions  and delay components to minimize the total bandwidth required by the packet delivery mechanism to ensure the E2E delay and overall packet loss probability, where routing and propagation delays are characterized by a deterministic backhaul delay, and the achievable rate in finite blocklength regime is applied. The E2E delay includes UL and DL transmission delay and queueing delay, the overall packet loss includes packet loss in UL and DL transmissions and queueing delay violation. A two-step method is proposed, where the bandwidth assignment is first optimized with given delay components and then the uplink and downlink bandwidth are optimized given the E2E delay. Numerical results show that the joint configuration requires half of the total bandwidth of the non-joint optimization.
\end{itemize}

The remainder of this paper is organized as follows. System model is described in Section II. A packet delivery mechanism is proposed in Section III. Section IV formulates the optimization problem. Section V shows how to obtain the optimal solution. Simulation and numerical results are provided in Section VI. Section VII concludes this work.

\section{System Model}
\subsection{Local Communication Scenarios}
We consider local communication scenarios, where BSs are connected with one-hop backhaul and the communication distance is less than a few kilometers. In such scenarios, propagation delay is negligible.  By deploying high-capacity backhaul links such as fiber, the backhaul delay is around $0.1$~ms \cite{Gongzheng2016Backhaul}.

\begin{figure}[htbp]
        \vspace{-0.3cm}
        \centering
        \begin{minipage}[t]{0.5\textwidth}
        \includegraphics[width=1\textwidth]{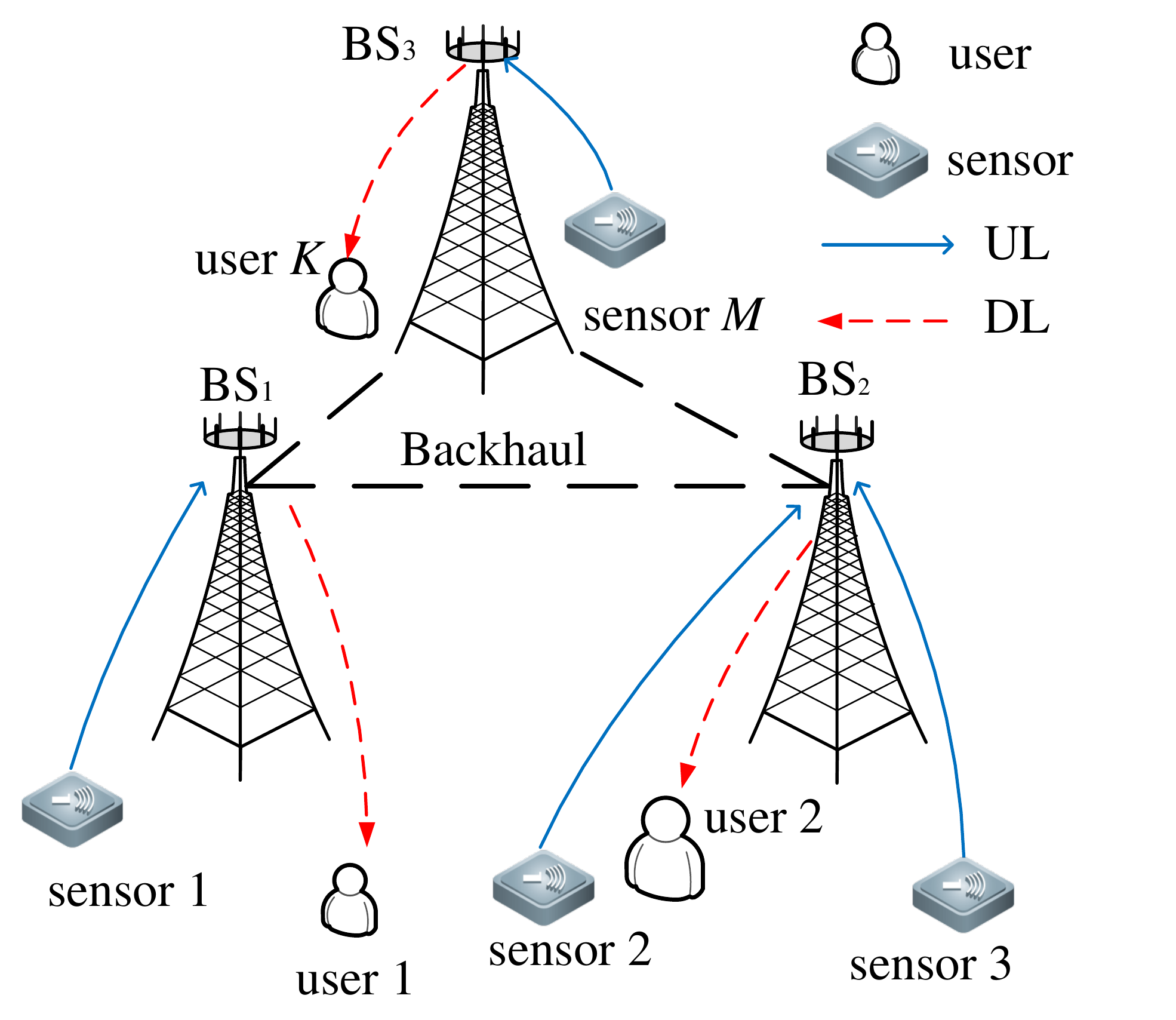}
        \end{minipage}
        \vspace{-0.2cm}
        \caption{Illustration of a local communication scenario.}
        \label{fig:system}
        \vspace{-0.6cm}
\end{figure}

As illustrated in Fig. \ref{fig:system}, we consider a cellular system with $K+M$ single-antenna devices and three BSs. Each device is served by one of the BSs, which are equipped with $N_\mathrm{t}$ antennas. To avoid strong interference among adjacent BSs, the frequency-reuse factor of the network is assumed to be $1/3$. Frequency-division multiple access is applied to avoid interference among different devices. There are two kinds of devices. The first kind of devices are $K$ users, which need to download packets from the BSs. The second kind of devices are $M$ sensors, which generate and upload packets to the BSs. After receiving the packets successfully, the BSs send the packets to users.\footnote{Device-to-device communications are possible for URLLC, which will not be addressed in this work.} Depending on application scenarios, each user may require packets from one or more sensors in the $M$ sensors, and the packets of one sensor may be required to one or more target users in the $K$ users. If a sensor and a target user are connected to two BSs, then the required packets need to be forwarded from the BS connected with the sensor to the BS connected with the user via backhaul. Frequency division duplex (FDD) systems is considered, because they are widely deployed. Our studies can be easily extended to time division duplex (TDD) systems. In TDD systems, one can adjust the ratio of UL and DL transmission durations, which is equivalent to adjusting UL and DL bandwidth in FDD systems.

\subsection{QoS Requirement} 
\begin{figure}[htbp]
        \vspace{-0.3cm}
        \centering
        \begin{minipage}[t]{0.6\textwidth}
        \includegraphics[width=1\textwidth]{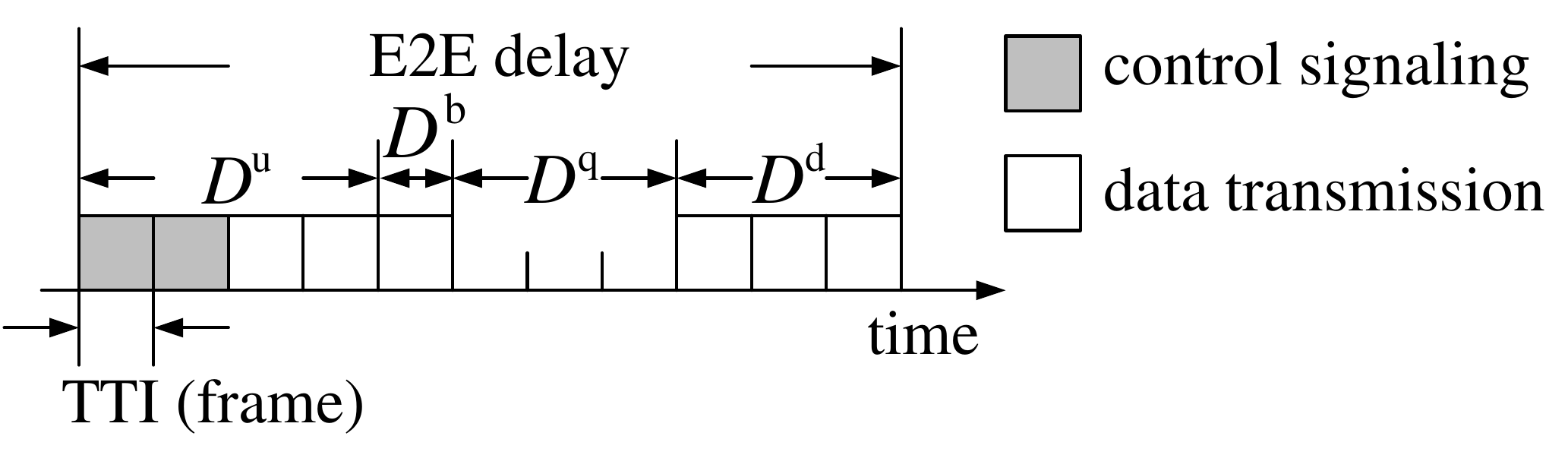}
        \end{minipage}
        \vspace{-0.2cm}
        \caption{Frame duration and delay components of event-driven packets.}
        \label{fig:time}
        \vspace{-0.2cm}
\end{figure}

The QoS requirement of URLLC is characterized by an ultra-short E2E delay $D_{\max}$ and an overall packet loss probability $\varepsilon_{\max}$ that are imposed on each packet. For local communication scenario, the E2E delay includes UL and DL transmission delays, queueing delay at the BSs and backhaul delay by assuming negligible propagation delay and processing delay. The overall packet loss comes from decoding errors and queueing delay violation. In Long Term Evolution systems, transmission time interval (TTI) is $1$~ms, and hence the E2E delay of URLLC cannot be supported. To reduce transmission delay, we consider the short frame structure as illustrated in Fig. \ref{fig:time}, where TTI equals to the frame duration $T_{\rm f}$, which is the minimal time granularity of the system (i.e., subframe in \cite{Shehzad2015Control}). Therefore, the transmission delays and queueing delay should be divisible by frame duration.
The E2E delay requirement is given by
\begin{align}
D^{\rm u} + D^{\rm d} + D^{\rm q} +D^{\rm b}\leq D_{\max},\label{eq:E2Ede}
\end{align}
where $D^{\rm u}$, $D^{\rm d}$, $D^{\rm q}$ and $D^{\rm b}$ are the UL and DL transmission delays, queueing delay and backhaul delay, respectively.

In application scenarios of URLLC such as smart factory and vehicle networks, there are two kinds of packets:  event-driven packets and periodic packets \cite{Shehzad2016Emerg,Hassan2013A}. For the event-driven packets, the transmission delay includes those caused by control signaling and data transmission  as illustrated in Fig. \ref{fig:time}. The UL transmission procedure includes the following steps: (i)  generation of a packet by a sensor; (ii) UL scheduling request from the sensor; (iii) bandwidth assignment and transmission grant by the BS; (iv) UL data transmission. The scheduling requests are triggered by some urgent events, and the BSs need to grant the transmission immediately when a request is received. To ensure that the UL scheduling request can be successfully received by the BS (i.e., to avoid scheduling collision), control channel should be reserved for each sensor with event-driven packets \cite{Shehzad2015Control}, and hence only two frames are occupied by control signaling in steps (ii) and (iii). For periodic packets, the inter-arrival time between packets are known at the BSs. By reserving data transmission resource to each sensor, there is no control signaling.\footnote{Since random access could cause long latency for machine-type communications \cite{Rana2017Random}, it will not be used for both kinds of packets in URLLC \cite{3GPP2016Scenarios}.}

Denote the requirements on the packet loss probabilities in UL transmission, DL transmission and queueing as ${\varepsilon ^{\rm u}}$, ${\varepsilon ^{\rm d}}$ and ${\varepsilon ^{\rm q}}$, respectively. The ultra-high reliability can be ensured if
\begin{align}
(1-{\varepsilon ^{\rm u}})(1 - {\varepsilon ^{\rm d}})(1-{\varepsilon ^{\rm q}})  \approx 1 - {\varepsilon ^{\rm u}}-{\varepsilon ^{\rm d}}-{\varepsilon ^{\rm q}} \leq {\varepsilon _{\max}}, \label{eq:E2Ere}
\end{align}
where the approximation is accurate since ${\varepsilon ^{\rm u}}$, ${\varepsilon ^{\rm d}}$, and ${\varepsilon ^{\rm q}}$ are extremely small.


\section{Packet Delivery Mechanism}
In this section, we propose a packet delivery mechanism for event-driven packets in URLLC, which can be easily extended to periodic packets since their arrival processes are deterministic. To reduce the total bandwidth required to ensure QoS requirement, we consider statistical multiplexing mode for queueing at BSs, broadcasting mode for downlink transmission, and bandwidth assignment for UL and DL transmissions.

\subsection{Queueing Mode}
As shown in \cite{3GPP2012MTC,Mehdi2013Performance}, the packet arrival process in vehicle networks and some M2M communications can be modeled as a Poisson process, which is an aggregation of packets generated by multiple sensors. Denote the average arrival rate of the Poisson process as $\lambda$~packets/frame. The arrival process of each sensor is modeled as Bernoulli process. Denote $a_m(n)$ as the number of packets arrived at a BS from the $m$th sensor in the $n$th frame, $m = 1,...,M$. With probability $p_m$, $a_m(n) = 1$, and with probability $1-p_m$, $a_m(n) = 0$. Then, the average total arrival rate of the $M$ sensors is $\lambda = \sum_{m=1}^Mp_m$~packets/frame. According to the result in \cite{She2017CrossLayer}, the effective bandwidth of the arrival process can be expressed as follows,
\begin{align}
E_{\rm B}  = \frac{T_{\rm f}\ln(1/\varepsilon^{\rm q})}{D^{\rm q} \ln \left[\frac{T_{\rm{f}}\ln(1/\varepsilon^{\rm q})}{\lambda D^{\rm q}}+1\right]}\;\text{packets/frame},\label{eq:sreq}
\end{align}
which is the minimal constant packet service rate required to ensure queueing delay $D^{\rm q}$ and queueing delay violation probability ${\varepsilon ^{\rm q}}$. It is widely believed that effective bandwidth is applicable when the queue length or queueing delay is long. However, the results in \cite{squeezing1996} imply that for Poisson process and the arrival processes that are more bursty than Poisson process, a short delay requirement $(D^{\rm q},{\varepsilon ^{\rm q}})$ can be satisfied with a constant packet service rate that is equal to or higher than $E_{\rm B}$ . This implication is validated in \cite{She2017CrossLayer} for typical arrival processes in URLLC, such as Poisson process, interrupted Poisson processes that is more bursty than Poisson process, and Switched Poisson process that is autocorrelated.

\begin{figure}[htbp]
        \vspace{-0.3cm}
        \centering
        \begin{minipage}[t]{0.3\textwidth}
        \includegraphics[width=1\textwidth]{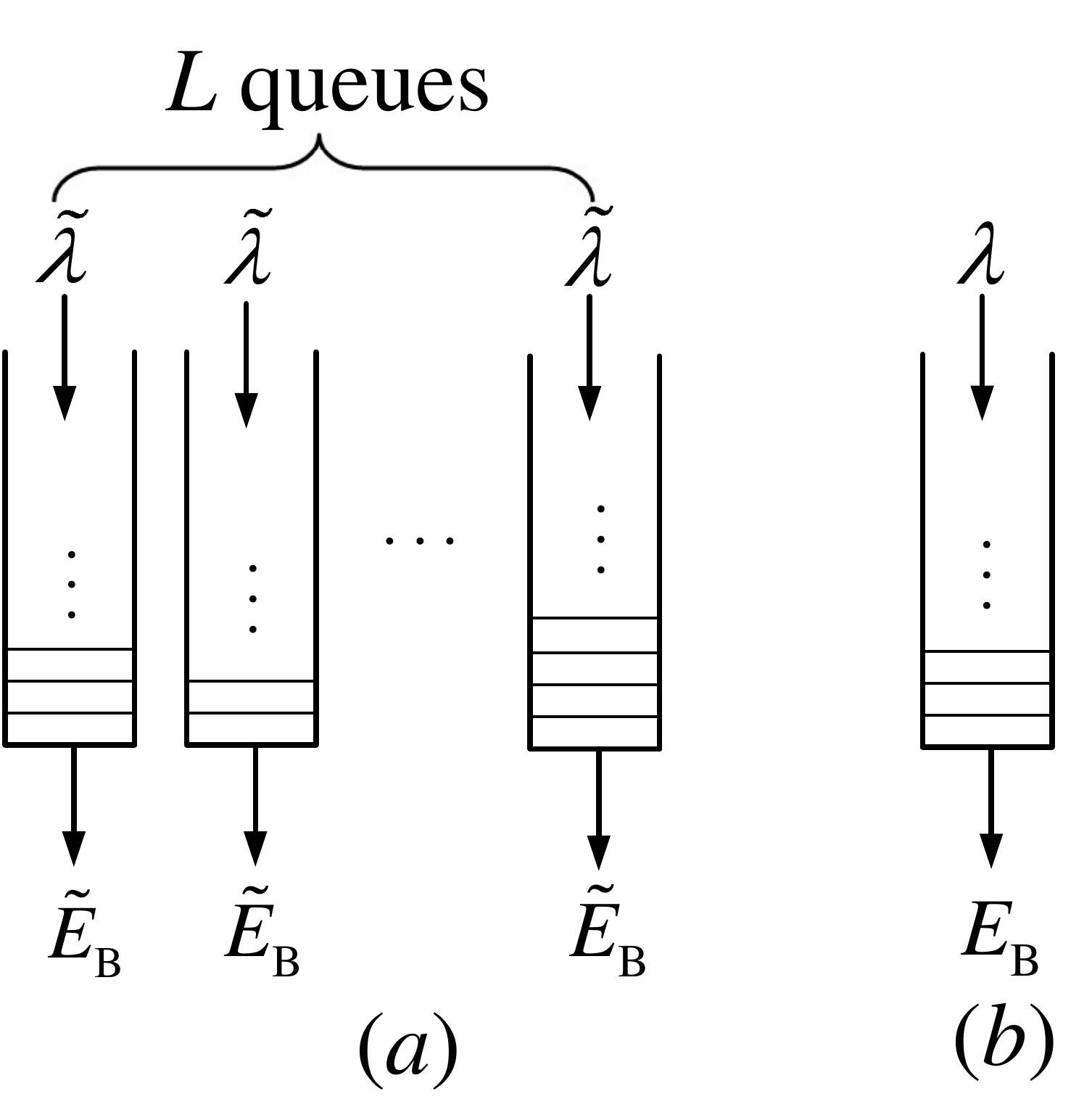}
        \end{minipage}
        \vspace{-0.2cm}
        \caption{Queueing modes. (a) Individual queueing mode. (b) Statistical multiplexing queueing mode.}
        \label{fig:Queue}
        \vspace{-0.4cm}
\end{figure}

In existing papers that study how to ensure queueing delay requirement of each users, the packets target to different users wait in different queues before DL transmission as in Fig. \ref{fig:Queue}(a)  \cite{She2017CrossLayer,Adnan2016Towards}. Such a queueing mode is referred to as \emph{individual queueing mode}. It has been shown that if $\tilde{\lambda} = \lambda/L$ and $\tilde{E}_{\rm B} = E_{\rm B}/L$, then the average queueing delay with the individual queueing mode will be $L$ times of that with the \emph{statistical multiplexing mode} in Fig. \ref{fig:Queue}(b) \cite{Mor2013Queue}. The following proposition indicates that a similar conclusion can be obtained when the delay requirement is characterized by $(D^{\rm q},{\varepsilon ^{\rm q}})$, which is imposed on each packet.

\begin{prop}\label{P:S}
\emph{Given the requirement on $D^{\rm q}$ and ${\varepsilon ^{\rm q}}$, if $\tilde{\lambda} = \lambda/L$, then $L\tilde{E}_{\rm B} > E_{\rm B}$.}
\begin{proof}
\emph{See proof in Appendix \ref{App:LS}.}
\end{proof}
\end{prop}

If the queueing delay requirement $(D_{\max},\varepsilon_q)$ can be satisfied with the statistical multiplexing queue, then for any packet that comes from any of the $M$ sensors, the probability that the queueing delay of the packet exceeds $D_{\max}$ is smaller than $\varepsilon_q$. In other words, no matter which sensor a packet came from, the delay requirement of it can be satisfied with the statistical multiplexing queue. Proposition \ref{P:S} indicates that to guarantee the queueing delay requirement imposed on each packet, the required effective bandwidth of statistical multiplexing queue is less than the total effective bandwidth of individual queues. Since effective bandwidth is the minimal constant service rate that can ensure queueing delay requirement, and the required bandwidth decreases with service rate, we consider statistical multiplexing mode for saving bandwidth.

\subsection{Transmission Mode and Bandwidth Assignment}
In typical scenarios of URLLC, the channel coherence time is longer than the typical E2E delay (i.e., $1$~ms) \cite{She2017CrossLayer}. Then, the channel gain changes little before the deadline of conveying each packet, and hence simply retransmitting a packet in multiple consecutive frames can hardly improve the reliability. When  channel is in deep fading, frequency diversity can be applied. To illustrate how the overall reliability can be ensured with diversity, we consider a simple method that transmits each packet multiple times over multiple separated subchannels \cite{David2014Achieving}.\footnote{For saving bandwidth, a simple idea is that all sensors transmit packet under the same spectrum. In the spectrum sharing network, interference is random and strong, which cannot be treated as additive noise or simply ignored. Unfortunately, the achievable rate with finite blocklength codes in the interference environment is unavailable in existing literatures. Therefore, whether or not the QoS requirement of URLLC can be satisfied in spectrum sharing systems is unknown and deserves further study.}


\subsubsection{Subchannel assignment for UL transmission} Considering that a sensor may stay dumb for a long duration between the transmissions of short packets \cite{3GPP2012MTC}, the bandwidth is only assigned to the active sensors immediately after receiving the scheduling requests.

\begin{figure}[htbp]
        \vspace{-0.3cm}
        \centering
        \begin{minipage}[t]{0.5\textwidth}
        \includegraphics[width=1\textwidth]{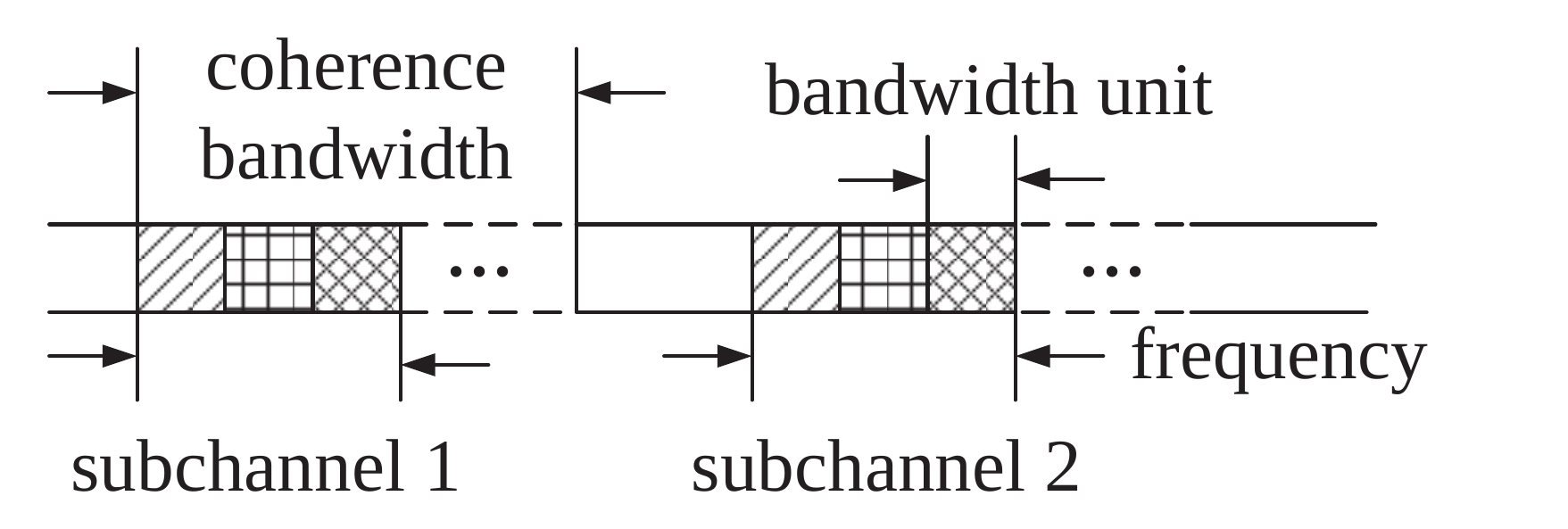}
        \end{minipage}
        \vspace{-0.2cm}
        \caption{Bandwidth of each subchannel.}
        \label{fig:frequency}
        \vspace{-0.2cm}
\end{figure}

To exploit frequency diversity, the BS assigns $N^{\rm u}_m$ separated subchannels to the $m$th sensor if it has a packet to transmit. The packet is repeatedly transmitted over the $N^{\rm u}_m$ subchannels. If one of the $N^{\rm u}_m$ transmissions is successful, then the packet is successfully received at the BS. Since the interference among sensors causes severe deterioration in QoS, we assume that different subchannels are assigned to the sensors that request transmissions concurrently. To maximize frequency diversity gain, the instantaneous channel gains on the $N^{\rm u}_m$ subchannels assigned to transmit one packet should be independent, i.e., the separation of the $N^{\rm u}_m$ subchannels should exceed the channel coherence bandwidth  $W_{\rm c}$, as  shown in Fig. \ref{fig:frequency}. In real-word systems, frequency is discretized into basic bandwidth units, e.g., subcarriers in orthogonal frequency division multiple access systems, and then each subchannel consists of multiple bandwidth units with bandwidth $B_0$. Denote $B^{\rm u}_m$ as the bandwidth of each subchannel allocated to the $m$th sensor. Then, $B^{\rm u}_m$ is divisible by $B_0$. By adjusting the number of bandwidth units in one subchannel, the bandwidth of each subchannel can be controlled. The bandwidth assigned to the sensor for transmitting a packet is $N^{\rm u}_m B^{\rm u}_m$. We assume that $B^{\rm u}_m<W_{\rm c}$, such that each subchannel is frequency-flat fading.

\subsubsection{Subchannel assignment for DL transmission}
We consider broadcast for DL transmission.\footnote{In some applications like factory automation and tactile internet each user only requests packets from $M'$ sensors, where $M' < M$. For these applications, multi-cast is an option for DL transmission, where users and sensors are grouped into multiple clusters. Our method can be extended into multi-cast systems by applying it in each cluster.} Without access control, acknowledgment feedback, and CSIT, the control and training/feedback overhead is negligible.
To guarantee reliability, each packet is repeatedly transmitted over $N^{\rm d}$ subchannels each with bandwidth $B^{\rm d} < W_{\rm c}$, and different packets are transmitted over different subchannels. Each user can receive the signals on all the DL subchannels. We assume that the channel coding on each subchannel is independent of the others. With independent channel coding, decoding errors on different subchannels are uncorrelated. If one packet is lost due to decoding error, other packets can still be decoded successfully.

\subsection{Illustration of the Packets Delivery Mechanism}
\begin{figure}[htbp]
        \vspace{-0.3cm}
        \centering
        \begin{minipage}[t]{0.55\textwidth}
        \includegraphics[width=1\textwidth]{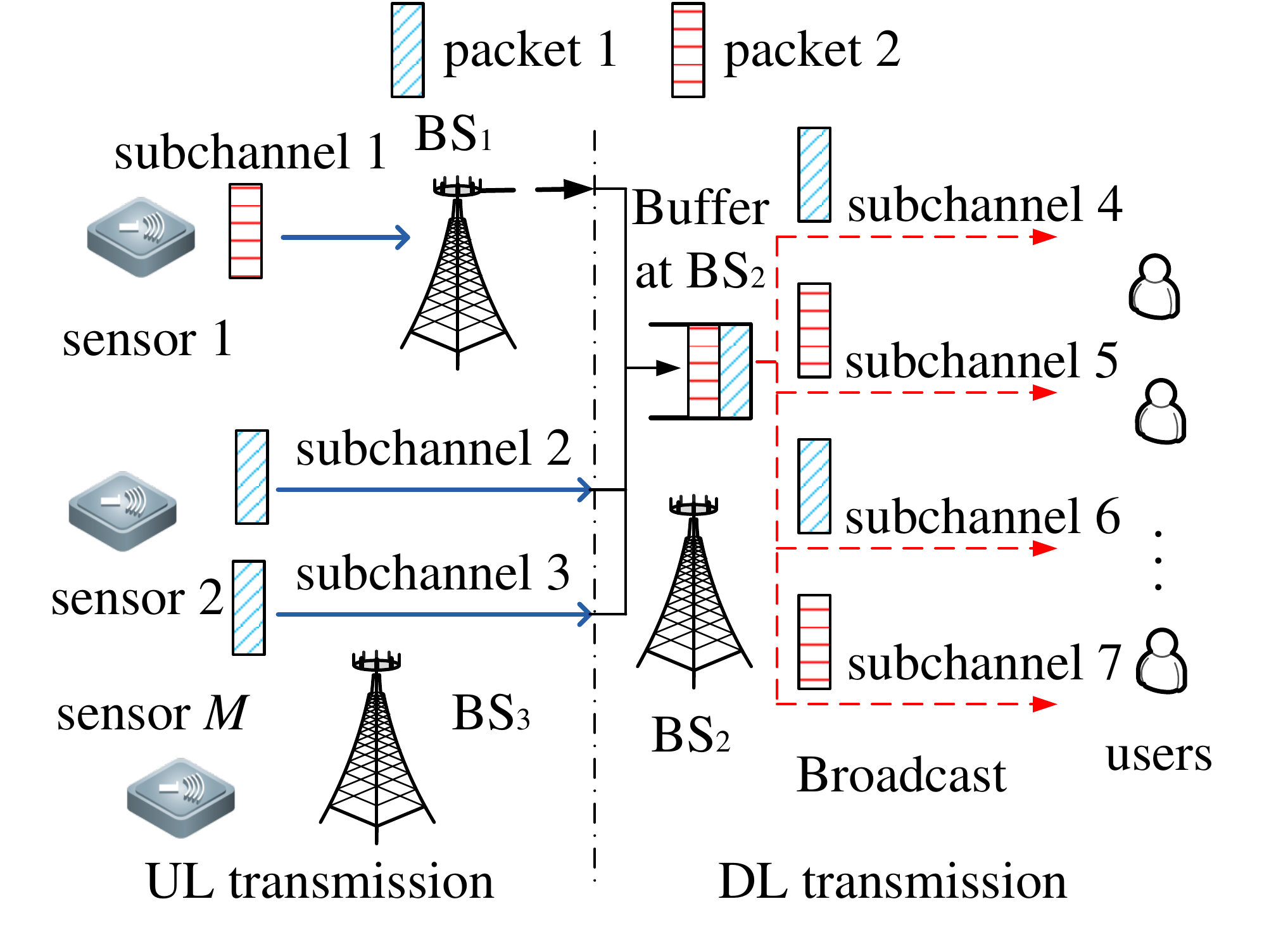}
        \end{minipage}
        \vspace{-0.2cm}
        \caption{Illustration of packets delivery mechanism.}
        \label{fig:mechanism}
        \vspace{-0.4cm}
\end{figure}

The proposed packets delivery mechanism is illustrated in Fig. \ref{fig:mechanism}, where sensors $1$, $2$, and $M$ are served by BSs $1$, $2$ and $3$, respectively. In the considered time slots, sensor $M$ has no packet to transmit and stays dumb, and the other two sensors are active. In the UL transmission phase, sensor $1$ sends a packet to $\text{BS}_1$ over subchannel $1$, and sensor $2$ sends two copies of its packet to $\text{BS}_2$ over subchannels $2$ and $3$. Then, $\text{BS}_1$ forwards the packet from sensor $1$ to $\text{BS}_2$ via backhaul. After arriving at the buffer of $\text{BS}_2$, both packets wait in the buffer before DL transmission. In the DL transmission phase, packets in the buffer are broadcast to all the users associated with $\text{BS}_2$ over multiple subchannels at rate $E_{\rm B}^+$~packets/frame, where ${E_{\rm B}^+}$ is the minimal integer that is equal to or higher than $E_{\rm B}$.

\section{Problem Formulation}
In this section, we formulate the optimization problem to minimize the total bandwidth by jointly optimizing UL and DL transmission delays, queueing delay and subchannel assignment.

\subsection{Ensuring UL and DL Packet Loss Requirements}
When analyzing the reliability of URLLC, Shannon capacity was applied in many existing studies such as \cite{Jie2017Availability,Beatriz2015Reliable}, which cannot characterize the decoding error probability.

\subsubsection{Constraint on UL transmission packet loss probability}
Denote the large-scale channel gain of the $m$th sensor as $\alpha^{\rm u}_m$, and the instantaneous channel gain on the $i$th subchannel allocated to the $m$th sensor as $g^{\rm u}_{m,i} = ({{\bf{h}}^{\rm u}_{m,i}})^H{\bf{h}}^{\rm u}_{m,i}$, where  $[ \cdot ]^H$ denotes the conjugate transpose and ${\bf{h}}^{\rm u}_{m,i} \in {\mathbb{C}}^{N_{\rm{t}} \times 1}$ is the channel vector whose elements are independent and identically complex Gaussian distributed with zero mean and unit variance. To avoid feedback overhead, CSIT is not assumed available at the sensors, and then the maximal transmit power at each sensor is equally allocated among $N^{\rm u}_m$ subchannels. For single-input-multiple-output system, the achievable rate from the $m$th sensor to the BS over the $i$th subchannel can be accurately approximated by \cite{Yury2014Quasi}
\begin{align}
R^{\rm{u}}_{m,i} \approx \frac{T_{\rm f}B^{\rm{u}}_m}{\ln{2}} \left\{\ln\left(1+\frac{\alpha^{\rm u}_m P^{\rm u}_{\max} g^{\rm u}_{m,i}}{ \phi N_0 B^{\rm{u}}_m N^{\rm{u}}_m}\right) - \sqrt{\frac{V^{\rm u}_{m,i}}{(D^{\rm u} - 2T_{\rm f}) B^{\rm{u}}_m}}f_{\rm Q}^{-1}({e^{\rm u}_{m,i} })\right\}\;\text{bits/frames}, \label{eq:snu}
\end{align}
where $P^{\rm u}_{\max}$ is the maximal transmit power of each sensor, $\phi > 1$ reflects the signal-to-noise ratio (SNR) loss due to the errors of channel estimation at receiver,\footnote{The impact of channel estimation errors on data rate can be equivalent to a SNR loss, which depends on the velocity of sensors \cite{Xin2015Energy}. Velocity of devices ranges from $0$ to $500$ km/h \cite{Philipp2017IoT}. For sensors with slow and median velocity, $\phi$ is close to $1$.}  $N_0$ is the single-sided noise spectral density, $e^{\rm u}_{m,i}$ is the decoding error probability (i.e.,
the block error probability) on the $i$th subchannel of the $m$th sensor, $f_{\mathrm Q}^{-1}(x)$ is the inverse of the Q-function, and $V^{\rm u}_{m,i} = 1-{\left[1+\frac{\alpha^{\rm u}_m P^{\rm u}_{\max} g^{\rm u}_{m,i}}{ {\phi}N_0 B^{\rm{u}}_m N^{\rm{u}}_m }\right]^{-2}}$ \cite{Yury2014Quasi}. The blocklength of channel coding is determined by the UL bandwidth of each subchannel and transmission duration according to $(D^{\rm u} - 2T_{\rm f}) B^{\rm u}_m$. When the blocklength is large, \eqref{eq:snu} approaches to the Shannon's capacity.

As shown in \cite{Yury2010Channel,Yury2014Quasi}, the approximation in \eqref{eq:snu} is very accurate in quasi-static channel when $e^{\rm u}_{m,i} \in [10^{-3},10^{-6}]$. In typical scenarios, the required transmission delay in URLLC is shorter than the channel coherence time \cite{She2017CrossLayer}, i.e., the channel is quasi-static.\footnote{If the transmission duration of each block is less than the channel coherence time, then the channel is referred to as quasi-static channel in \cite{Yury2014Quasi}.} This suggests that the approximation in \eqref{eq:snu} is applicable for URLLC.

When transmitting a packet that contains $b$ bits over one subchannel, the decoding error probability can be obtained by substituting \eqref{eq:snu} into $R^{\rm u}_{m,i}(D^{\rm u} - 2T_{\rm f})/T_{\rm f} = b$ as
\begin{align}
e^{\rm u}_{m,i} = {f_{\rm{Q}}}\left\{ {\sqrt {{{({D^{\rm{u}}} - 2{T_{\rm{f}}})B_m^{\rm{u}}}}} \left[ {\ln \left( {1 + \frac{{\alpha _m^{\rm{u}}P_{\max }^{\rm{u}}g_{m,i}^{\rm{u}}}}{{{{\phi}N_0}B_m^{\rm{u}}N_m^{\rm{u}}}}} \right) - \frac{{b\ln 2}}{{B_m^{\rm{u}}({D^{\rm{u}}} - 2{T_{\rm{f}}})}}} \right]} \right\},\label{eq:TrE}
\end{align}
where $V^{\rm u}_m \approx 1$ is applied. $e^{\rm u}_{m,i}$ in \eqref{eq:TrE} depends on channel, and hence is a random variable. To show the relation between the decoding error probability and packet loss probability, we use indicator functions to represent whether a packet is successfully transmitted over multiple subchannels. If the packet is successfully transmitted to the BS over the $i$th subchannel assigned to the $m$th sensor, then ${\bf{1}}_{m,i}^{\rm u} = 1$. Otherwise, ${\bf{1}}_{m,i}^{\rm u} = 0$. From \eqref{eq:TrE}, we have
\begin{align}
\Pr\{{\bf{1}}_{m,i}^{\rm u} = 0\} = \int_0^\infty  {e_{m,i}^{\rm{u}}{f_g}\left( x \right)} dx, \label{eq:Prsuc}
\end{align}
where ${f_{\rm{g}}}\left( x \right)$ is the distribution of instantaneous channel gain. If the elements of ${\bf{h}}^{\rm u}_{m,i}$ are complex Gaussian distributed, then $f_{\rm g}\left( x \right) = \frac{1}{{\left( {{N_\mathrm{t}} - 1} \right)!}}{x^{{N_\mathrm{t}} - 1}}{e^{ - x}}$. Since each packet is transmitted over $N_m^{\rm u}$ subchannels, the packet loss probability is given by
\begin{align}
\Pr \left\{ \mathop  \cap \limits_{i = 1}^{N_m^{\rm u}} \left( {{\bf{1}}_{m,i}^{\rm u} = 0} \right)\right\} = \prod\limits_{i = 1}^{N^{\rm u}_m} {\Pr \left\{ {\bf{1}}_{m,i}^{\rm u} = 0 \right\}} = \left[\int_0^\infty  {e_{m,i}^{\rm{u}}{f_g}\left( x \right)} dx\right]^{N_m^{\rm u}}\label{eq:ULloss},
\end{align}
which should be no large than ${\varepsilon ^{\rm u}}$ to guarantee the UL reliability.

Optimizing resource allocation under the constraint on the packet loss probability in \eqref{eq:ULloss} is very difficult, because the expression of $e_{m,i}^{\rm{u}}$ in \eqref{eq:TrE} is too complicated to obtain any useful insights. To simplify the analysis, we consider an upper bound of it. Since $f_{\rm Q}(\cdot)$ is a decreasing function, $e^{\rm u}_{m,i}$ decreases as $g_{m,i}^{\rm{u}}$ increases. Then, an upper bound of $e^{\rm u}_{m,i}$ can be obtained from
\begin{align}
e_{m,i}^{\rm{u}} \le \left\{ {\begin{array}{*{20}{c}}
   {e_m^{{\rm{u,th}}},{\rm{if}}\;g_{m,i}^{\rm{u}} \ge g_m^{{\rm{u,th}}},}  \\
   {1,\quad\;{\rm{if}}\;g_{m,i}^{\rm{u}} < g_m^{{\rm{u,th}}},}  \\
\end{array}} \right.\label{eq:UBe}
\end{align}
where $g_m^{{\rm{u,th}}}$ can be obtained by substituting $e_m^{{\rm{u,th}}}$ into $R^{\rm u}_{m,i}(D^{\rm u} - 2T_{\rm f})/T_{\rm f} = b$ as
\begin{align}
g_m^{\rm{u,th}} \approx \frac{{{{\phi}N_0}{B^{\rm u}_m}N^{\rm u}_m}}{{{\alpha^{\rm u} _m}{P^{\rm u}_{\max}}}}\left\{ {\exp \left[ {\frac{{b\ln 2}}{{(D^{\rm u} - 2T_{\rm f}) {B^{\rm u}_m}}} + \sqrt {\frac{1}{{(D^{\rm u} - 2T_{\rm f}) {B^{\rm u}_m}}}} f_{\rm Q}^{ - 1}\left( {e^{\rm u,th}_{m} } \right)} \right] - 1} \right\}.\label{eq:thresh}
\end{align}
The upper bound in \eqref{eq:UBe} means that if the instantaneous channel gain $g^{\rm u}_{m,i}$ is higher than a threshold $g_m^{\rm{u,th}}$ such that $R^{\rm u}_{m,i}(D^{\rm u} - 2T_{\rm f})/T_{\rm f} \geq b$, then a packet with size $b$ can be transmitted successfully with probability $1 - e^{\rm u, th}_m$ over the $i$th subchannel. Otherwise, the packet cannot be transmitted successfully over the $i$th subchannel.

The upper bound of decoding error probability in \eqref{eq:UBe} is different from outage probability, defined as the probability that the SNR or SINR is lower than a threshold in \cite{Jie2017Availability,Beatriz2015Reliable}. When the channel gain exceeds the threshold, the outage probability is zero. As shown in  \eqref{eq:UBe}, however, even when the channel gain is higher than the threshold $g_m^{\rm{u,th}}$, ${e^{\rm u,th}_{m} }$ is not zero. The relation between ${e^{\rm u,th}_{m} }$ and $g_m^{\rm{u,th}}$ is shown in \eqref{eq:thresh}.

From \eqref{eq:UBe}, the packet loss probability can be bounded by
\begin{align}
\Pr \left\{ \mathop  \cap \limits_{i = 1}^{N_m^{\rm u}} \left( {{\bf{1}}_{m,i}^{\rm u} = 0} \right)\right\} \leq {\left(  \Pr \left\{ g^{\rm u}_{m,i} < g^{\rm th, u}_m \right\} + e^{\rm u,th}_m \right)^{N^{\rm u}_m}}\label{eq:Prg}.
\end{align}
Since $\Pr \left\{ g^{\rm u}_{m,i} < g^{\rm th, u}_m \right\} = \int_0^{g^{\rm{u,th}}_m} {{f_{\rm{g}}}\left( x \right)dx}$, the constraint on UL packet loss probability is
\begin{align}
f_m^{\rm u}(N^{\rm u}_m,B^{\rm u}_m, e^{\rm u,th}_{m}) \triangleq {\left[ {\int_0^{g^{\rm{u,th}}_m} {{f_{\rm{g}}}\left( x \right)dx} } + e^{\rm u,th}_m \right]^{N^{\rm u}_m}}   \leq \varepsilon^{\mathrm u}, m=1,...,M.\label{eq:ULTr}
\end{align}

\begin{rem}\label{R:gth}
\emph{The upper bound in \eqref{eq:UBe} is not accurate when $g^{\rm u}_{m,i}$ is smaller than the threshold $g_m^{\rm{u,th}}$. In this case, the decoding error probability ranges from $e^{\rm u, th}_m$ to $1$. Since
	$e^{\rm u, th}_m$ is smaller than $1$, setting $e_{m,i}^{\rm{u}} = 1$ when $g^{\rm u}_{m,i} < g_m^{\rm{u,th}}$ leads to conservative resource allocation. We will show the impact of the loose upper bound on bandwidth allocation with numerical results.}
\end{rem}

\subsubsection{Constraint on DL transmission packet loss probability} Since the frame duration is much shorter than the E2E delay, it is possible to adjust transmission duration of each packet $D^{\rm d}$. To achieve a constant rate of ${E_{\rm B}^+}$ packets per frame, the number of packets that are transmitted simultaneously is $\frac{D^{\rm d}}{T_{\rm f}}{E_{\rm B}^+}$. Without CSIT, the maximal transmit power of a BS $P^{\rm d}_{\max}$ is equally allocated among $\frac{D^{\rm d}}{T_{\rm f}}{E_{\rm B}^+}N^{\rm d}$ active subchannels. Denote the average channel gain of the $k$th user as $\alpha^{\rm d}_k$.



Similar to UL transmission, we can derive the threshold in DL transmission as follows:
\begin{align}
g_k^{\rm{d,th}} \approx \frac{{{{\phi}N_0}{B^{\rm d}}{D^{\rm d}}{E_{\rm B}^+}N^{\rm d} N_{\rm t}}}{{{\alpha^{\rm d} _k}{P^{\rm d}_{\max}}{T_{\rm f}}}}\left\{ {\exp \left[ {\frac{{b\ln 2}}{{D^{\rm d} {B^{\rm d}}}} + \sqrt {\frac{1}{{D^{\rm d} {B^{\rm d}}}}} f_{\rm Q}^{ - 1}\left( {e^{\rm d,th}_{k} } \right)} \right] - 1} \right\},\label{eq:threshDL}
\end{align}
where $e^{\rm d,th}_{k}$ is the block error probability when the instantaneous channel gain is $g_k^{\rm{d,th}}$. Then, the probability that the packet is not successfully transmitted to the $k$th user is bounded by
\begin{align}
f_k^{\rm d}(N^{\rm d},B^{\rm d}, e^{\rm d,th}_k) \triangleq {\left[ {\int_0^{g^{\rm{d,th}}_k} {{f_{\rm{g}}}\left( x \right)dx} } + e^{\rm d,th}_k \right]^{N^{\rm d}}}\label{eq:DLTR1}.
\end{align}
As shown in \eqref{eq:threshDL}, $g^{\rm{d,th}}_k$ decreases with $\alpha^{\rm d}_k$. Moreover, $f_k^{\rm d}(N^{\rm d},B^{\rm d}, e^{\rm d,th}_k)$ increases with $g^{\rm{d,th}}_k$. Therefore, the user with the lowest average channel gain has the highest packet loss probability. To study DL transmission reliability for all users, we only need to consider the user with the index ${k_{\min}} = \mathop {\arg }\limits_k \min \alpha _k^{\rm{d}}$. Then, the constraint on DL packet loss probability is,
\begin{align}
f_{k_{\min}}^{\rm d}(N^{\rm d},B^{\rm d}, e^{\rm d,th}_{k_{\min}}) \leq \varepsilon^{\mathrm d} \label{eq:DLTR}.
\end{align}

\begin{rem}\label{R:V}
\emph{Since $V^{\rm u}_{m,i} \leq 1$, by substituting $V^{\rm u}_{m,i}=1$ into \eqref{eq:snu}, we can obtain a lower bound of the achievable rate. As validated in \cite{Gross2015Delay}, $V$ is very close to $1$ when the SNR is higher than $10$~dB, which is the typical SNR at the edge of a cell \cite{David2005Fundamentals}. On the other hand, to guarantee the QoS requirement of typical applications in URLLC \cite{Philipp2017IoT}, the required SNR should be high, which can be supported by equipping multiple antennas at the BS. To simplify the analysis, we use the lower bounds in the rest of this work, i.e., $V^{\rm u}_{m,i}=1$. For DL transmission, we can also obtain a lower bound of achievable rate in this way.}
\end{rem}

\subsection{Total Bandwidth of the System}
To ensure $D^{\rm q}$ and ${\varepsilon ^{\rm q}}$, the packet rate for DL transmission should be ${E_{\rm B}^+}$ packets per frame. To ensure downlink transmission delay of each packet $D^{\rm d}$ with frame duration $T_{\rm f}$, the number of packets that are transmitted simultaneously should be $\frac{D^{\rm d}}{T_{\rm f}}{E_{\rm B}^+}$. Therefore, the bandwidth required for DL transmission in each cell is $\frac{D^{\rm d}}{T_{\rm f}}{E_{\rm B}^+}N^{\rm d}B^{\rm d}$. Denote the number of active sensors in one frame as $M_{\rm a}$, and the indices of these sensors as set ${\mathcal{M}}_{\rm a}$. Then, the total bandwidth for UL transmission is $\sum\limits_{m \in {\mathcal{M}}_{\rm a}} {N^{\rm u}_m}{B^{\rm u}_m}$. Thus, the required total bandwidth of the system is given by
\begin{align}
\sum\limits_{m \in {\mathcal{M}}_{\rm a}} {N^{\rm u}_m}{B^{\rm u}_m} + F_{\rm R}^{-1}\frac{D^{\rm d}}{T_{\rm f}}{{E_{\rm B}^+}}{N^{\rm d}}{B^{\rm d}}\label{eq:ob1},
\end{align}
where $F_{\rm R}$ is the frequency-reuse factor.

Since $M_{\rm a}$ is a random variable, and the resource allocation changes with $M_{\rm a}$, the BSs need to solve the optimization problem when the number of active sensors changes (e.g., every millisecond). To reduce computational complexity for solving optimization problem, we introduced an upper bound of the number of active sensors.

Denote ${\bf{1}}_m$ as an indicator function. If the $m$th sensor is active, ${\bf{1}}_m = 1$. Otherwise, ${\bf{1}}_m = 0$. Then, ${\mathbb{E}}({\bf{1}}_m)$ can be expressed as a function of $p_m$, which is the probability that the $m$th sensor has a transmission request in each frame. In particular, if there is a request, the sensor will stay active in $({D^{\rm{u}}-2T_{\rm f}})/{T_{\rm f}}$ frames, and ${\mathbb{E}}({\bf{1}}_m) = ({D^{\rm{u}}-2T_{\rm f}})p_m/{T_{\rm f}}$. Since all the sensors could be active at the same time, a simple upper bound of \eqref{eq:ob1} can be obtained by setting $M_{\rm a} = M$. However, such an bound will lead to very conservative bandwidth assignment for UL transmission if the number of sensors is large, where the probability that all the sensors are active is extremely small. In what follows, we provide a threshold of $M_{\rm a}$, which is an upper bound of $M_{\rm a}$ with high probability. Denote the threshold as $M_{\rm a}^{\rm th}$. With probability ${\varepsilon _{\rm M}}$, $M_{\rm a}$ is higher than $M_{\rm a}^{\rm th}$, i.e., ${\varepsilon _{\rm M}} \triangleq \Pr\{M_{\rm a} > M_{\rm a}^{\rm th}\}$. Since $M_{\rm a}$ is the sum of $M$ Bernoulli process, it can be approximated as a Poisson process with parameter $\frac{D^{\rm{u}}-2T_{\rm f}}{T_{\rm f}}\sum_{m=1}^Mp_m$. Hence, it is not hard to obtain $M_{\rm a}^{\rm th}$ with given ${\varepsilon _{\rm M}}$. Then, the bandwidth for UL transmission is bounded by $\frac{M_{\rm a}^{\rm th}}{M}\sum\limits_{m = 1}^M N_m^{\rm u} B_m^{\rm u}$ with high probability. When $M_{\rm a} > M_{\rm a}^{\rm th}$, some packets may be lost due to insufficient bandwidth, which however has little impact on the overall reliability if ${\varepsilon _{\rm M}} \ll \varepsilon_{\max}$. Then, an upper bound of the required total bandwidth of the system can be obtained as $\frac{M_{\rm a}^{\rm th}}{M}\sum\limits_{m = 1}^M N_m^{\rm u} B_m^{\rm u}+F^{-1}_{\rm R}\frac{D^{\rm d}}{T_{\rm f}}{{E_{\rm B}^+}}{N^{\rm d}}{B^{\rm d}} \triangleq B_{\rm tot}$. With this upper bound, the BSs only need to solve the optimization problem and update resource allocation when the large-scale channel gains change (e.g., every second).

\subsection{Optimization Problem}
The optimal UL and DL transmission delays, queueing delay and subchannel assignment that minimize the upper bound of the required total bandwidth to ensure the QoS can be obtained from the following problem:
\begin{align}
\mathop {\mathop {\min }\limits_{{D^{\rm{u}}},{D^{\rm{d}}},{D^{\rm{q}}},{N^{\rm d}}{B^{\rm d}},e^{\rm d,th}_{k_{\min}}} }\limits_{N_m^{\rm u},B_m^{\rm u},e^{\rm u,th}_{m},m = 1,...,M} & \frac{M_{\rm a}^{\rm th}}{M}\sum\limits_{m = 1}^M N_m^{\rm u} B_m^{\rm u}+F^{-1}_{\rm R}\frac{D^{\rm d}}{T_{\rm f}}{{E_{\rm B}^+}}{N^{\rm d}}{B^{\rm d}}\label{eq:aveBand}\\
\text{s.t.} \; &{E_{\rm B}^+} = \left\{\frac{\ln(1/\varepsilon^{\rm q})}{D^{\rm q} \ln \left[\frac{T_{\rm{f}}\ln(1/\varepsilon^{\rm q})}{\lambda D^{\rm q}}+1\right]}\right\}^+,\tag{\theequation a}\label{eq:constantS} \\
& 0 < {B^{\rm u}_m} \le {W_{\rm c}},0 < {B^{\rm d}} \le {W_{\rm c}},{B^{\rm u}_m},{B^{\rm d}} \in \{zB_0, z \in {\mathbb{Z}}\}\label{eq:Wcall}\tag{\theequation b} \\
& 0 < N^{\rm u}_m, 0 < N^{\rm d}, N^{\rm u}_m,N^{\rm d} \in {\mathbb{Z}}, \label{eq:Nall}\tag{\theequation c}\\
& D^{\rm{u}} \in \{ 3T_{\rm f},4T_{\rm f},...,D_{\max}-D^{\rm b}-2T_{\rm f}\},\nonumber\\
& D^{\rm{d}}, D^{\rm{q}} \in \{T_{\rm f},2T_{\rm f},...,D_{\max}-D^{\rm b} 4T_{\rm f} \},\nonumber\\
& \eqref{eq:E2Ede}, \eqref{eq:ULTr},\;\text{and}\;\eqref{eq:DLTR},\nonumber
\end{align}
where constraint in \eqref{eq:constantS} is the required DL packet service rate for ensuring queueing delay and queueing delay bound violation probability, constraint \eqref{eq:Wcall} ensures the bandwidth of each subchannel less than the coherence bandwidth such that each copy of a packet is transmitted over a flat fading channel,\footnote{The value of $W_{\rm c}$ depends on propagation environment, which is not hard to obtain before a system is configured \cite{3GPP2010Shadowing}.} \eqref{eq:E2Ede} is the constraint on the E2E delay, and the constraints in \eqref{eq:ULTr} and \eqref{eq:DLTR} ensure the transmission packet loss probabilities in UL and DL, respectively. Because the upper bounds in \eqref{eq:ULTr} and \eqref{eq:DLTR} are not unique and depend on $e^{\rm u, th}_m$ and $e^{\rm d,th}_{k_{\min}}$, the values of $e^{\rm u, th}_m$ and $e^{\rm d,th}_{k_{\min}}$ affect the optimal solution and the total bandwidth. To minimize the required total bandwidth, we adjust $e^{\rm u, th}_m$ and $e^{\rm d,th}_{k_{\min}}$ in the upper bounds in an optimal manner.

\begin{rem}\label{R:epson}
\emph{Similar to the delay components, the system can also adjust packet loss components in queueing and UL and DL transmissions. With different values of $\varepsilon^{\mathrm u}$, $\varepsilon^{\mathrm q}$ and $\varepsilon^{\mathrm d}$, the required total bandwidth is different. In problem \eqref{eq:aveBand}, the values of $\varepsilon^{\mathrm u}$, $\varepsilon^{\mathrm q}$ and $\varepsilon^{\mathrm d}$ are given. If they are optimization variables, then problem \eqref{eq:aveBand} will become intractable. Owing to the following reason, we can provide a simple but reasonable way to divide the constraint on the overall packet loss probability into constraints on different components. If $\varepsilon^{\mathrm u} \to 0$, then according to \eqref{eq:ULTr}, $N^{\rm u}_m \to \infty$, which means that the required bandwidth tends to infinite. Similarly, if $\varepsilon^{\mathrm d} \to 0$, $N^{\rm d} \to \infty$. Moreover, by substituting \eqref{eq:sreq} into \eqref{eq:ob1}, the relation between $B_{\rm tot}$ and $\varepsilon^{\mathrm q}$ can be expressed as ${B_{{\rm{tot}}}} = {C_1} + {C_2}{{\ln \left( {1/{\varepsilon ^{\rm{q}}}} \right)}}/{{\ln \left[ {{C_3}\ln \left( {1/{\varepsilon ^{\rm{q}}}} \right) + 1} \right]}}$, where $C_1$, $C_2$ and $C_3$ are parameters that do not change with $\varepsilon ^{\rm{q}}$. If $\varepsilon^{\mathrm q} \to 0$, then ${B_{{\rm{tot}}}}$ approaches infinite. This suggests that all the values of $\varepsilon^{\mathrm u}$, $\varepsilon^{\mathrm q}$ and $\varepsilon^{\mathrm d}$ cannot be ignored. Here we set them as one third of $\varepsilon_{\max}$, i.e., $\varepsilon^{\mathrm u} = \varepsilon^{\mathrm q} = \varepsilon^{\mathrm d} = \varepsilon_{\max}/3$. In simulation part, we will show that the total bandwidth with optimal values of $\varepsilon^{\mathrm u}$, $\varepsilon^{\mathrm q}$ and $\varepsilon^{\mathrm d}$ is almost the same as that with $\varepsilon^{\mathrm u} = \varepsilon^{\mathrm q} = \varepsilon^{\mathrm d} = \varepsilon_{\max}/3$.}
\end{rem}

\section{Joint UL and DL Resource Configuration}
In this section, we provide a two-step method to find the optimal solution of problem \eqref{eq:aveBand}. In the first step, we find the optimal subchannel assignment that minimizes total bandwidth with give delay components. In the second step, we find the optimal delay components that minimize the total bandwidth. Finally, we prove that the two-step method can provide the optimal solution.

\subsection{Bandwidth Assignment Optimization}
In this subsection, we fix the values of ${D^{\rm u}}$, ${D^{\rm d}}$ and ${D^{\rm q}}$, and optimize the values of $N^{\rm u}_m$, $B^{\rm u}_m$, $N^{\rm d}$ and $B^{\rm d}$. Because the UL and DL bandwidth assignments can be decoupled given the delay components and packet loss components, we first optimize UL subchannel assignment, and then consider DL subchannel assignment.

\subsubsection{UL subchannel assignment}
We optimize the values of $N^{\rm u}_m$, $B^{\rm u}_m$, and $e^{\rm u,th}_m$ in \eqref{eq:UBe} to minimize the UL bandwidth under the constraint on $\varepsilon^{\mathrm u}$ with given  ${D^{\rm u}}$ from the following problem:
\begin{align}
\mathop {\mathop {\min}\limits_{N^{\rm u}_m,{B^{\rm u}_m},e^{\rm u,th}_m } }\limits_{m = 1,...,{M}} & \frac{M_{\rm a}^{\rm th}}{M}\sum\limits_{m = 1}^M N_m^{\rm u} B_m^{\rm u} \label{eq:Wtot}\\
\text{s.t.}\;& 0 < {B^{\rm u}_m} \le {W_{\rm c}}, 0 < N^{\rm u}_m, N_m \in {\mathbb{Z}},\label{eq:Wc}\tag{\theequation a} \\
& \text{and}\;\eqref{eq:ULTr},\nonumber
\end{align}
where $B_m^{\rm u}$ is relaxed to continuous variable. After obtaining the solution of problem \eqref{eq:Wtot}, the discrete value of $B_m^{\rm u}$ can be directly obtained from $(B_m^{\rm u}/B_0)^+B_0$.

In the sequel, we propose an algorithm to find the optimal solution of problem \eqref{eq:Wtot}. Since the constraints for each sensor do not depend on those of the other sensors, problem \eqref{eq:Wtot} can be further equivalently decomposed into $M$ single-sensor problems as follows:
\begin{align}
\mathop {\mathop {\min}\limits_{N^{\rm u}_m,{B^{\rm u}_m},e^{\rm u,th}_m } } & {N^{\rm u}_m{B^{\rm u}_m}} \label{eq:su}\\
\text{s.t.} & \;\eqref{eq:Wc}, \;\text{and}\;\eqref{eq:ULTr}.\nonumber
\end{align}

To solve problem \eqref{eq:su}, we need some properties of $f_m^{\rm u}(N^{\rm u}_m,B^{\rm u}_m, e^{\rm u,th}_m)$.
\begin{pro}\label{P:fw}
\emph{Given the values of $N^{\rm u}_m$ and $e^{\rm u,th}_m$, we can find a unique solution of ${B}^{\rm u,min}_m$ that minimizes $f_m^{\rm u}(N^{\rm u}_m,B^{\rm u}_m, e^{\rm u,th}_m)$. Moreover, $f_m^{\rm u}(N^{\rm u}_m,B^{\rm u}_m, e^{\rm u,th}_m)$ strictly decreases with $B^{\rm u}_m$ in the region $B^{\rm u}_m \in [0,{B}^{\rm u,mim}_m]$ and strictly increases with $B^{\rm u}_m$ in the region $B^{\rm u}_m \in [{B}^{\rm u,mim}_m,\infty)$.}
\begin{proof}
\emph{See proof in Appendix \ref{App:Appendix_P1}.}
\end{proof}
\end{pro}

According to the numerical results in \cite{Chengjian2017GC}, ${B}^{\rm u,min}_m$ is larger than $W_{\rm c}$ in typical scenarios of URLLC. In what follows, we propose an algorithm that can find the global optimal solution when ${B}^{\rm u,min}_m  \geq W_{\rm c}$. For the case ${B}^{\rm u,min}_m < W_{\rm c}$, a local optimal solution can be obtained.

Based on Property \ref{P:fw}, we can obtain the following property.
\begin{pro}\label{P:w}
\emph{Given the value of $N^{\rm u}_m$, when ${B}^{\rm u,min}_m  \geq W_{\rm c}$, $f_m^{\rm u}(N^{\rm u}_m,B^{\rm u}_m, e^{\rm u,th*}_m)$ strictly decreases with $B^{\rm u}_m$ in the region $[0,W_{\rm c}]$, where $e^{\rm u,th*}_m$ is the optimal value that minimizes $f_m^{\rm u}(N^{\rm u}_m,B^{\rm u}_m, e^{\rm u,th}_m)$ with given $N^{\rm u}_m$ and $B^{\rm u}_m$.}
\begin{proof}
\emph{See proof in Appendix \ref{App:Appendix_P2}.}
\end{proof}
\end{pro}

If $f_m^{\rm u}(N^{\rm u}_m,W_{\rm c}, e^{\rm u,th*}_m) > \varepsilon^{\mathrm u}$, the reliability can not be guaranteed, and more subchannels are needed.
If $f_m^{\rm u}(N^{\rm u}_m,W_{\rm c}, e^{\rm u,th*}_m) \leq \varepsilon^{\mathrm u}$, the minimal value of $B^{\rm u}_m$ that satisfies \eqref{eq:ULTr} can be obtained when $f_m^{\rm u}(N^{\rm u}_m,B^{\rm u}_m, e^{\rm u,th*}_m) = \varepsilon ^{\mathrm u}$, and can be obtained via the binary search method \cite{boyd}. The search algorithm needs to compute the value of $f_m^{\rm u}(N^{\rm u}_m,B^{\rm u}_m, e^{\rm u,th*}_m)$, and hence needs to find $e^{\rm u,th*}_m$ with given $B^{\rm u}_m$. To show when $e^{\rm u,th*}_m$ can be obtained with a low complexity method, we provide the following property.

\begin{pro}\label{P:fe}
\emph{Given the values of $N^{\rm u}_m$ and $B^{\rm u}_m$, $f_m^{\rm u}(N^{\rm u}_m,B^{\rm u}_m, e^{\rm u,th}_m)$ is convex in $e^{\rm u,th}_m$ when $g^{\rm u,th}_m < N_{\rm t}-1$.}
\begin{proof}
\emph{See proof in Appendix \ref{App:Appendix_P3}.}
\end{proof}
\end{pro}

If $f_m^{\rm u}(N^{\rm u}_m,B^{\rm u}_m, e^{\rm u,th}_m)$ is convex in $e^{\rm u,th}_m$, then $e^{\rm u,th*}_m$ can be obtained by the exact linear search method \cite{boyd}. Otherwise, to obtain $e^{\rm u,th*}_m$, the exhaustive search method should be used. Note that to ensure ultra-high reliability in \eqref{eq:ULTr}, $g^{\rm u,th}_m$ cannot be too large. For example, when $N_{\rm t} \geq 2$ and $\varepsilon_{\max} \leq 10^{-5}$ ( which is typical for URLLC), we have $g^{\rm u,th}_m < N_{\rm t}-1$ under constraint \eqref{eq:ULTr} in the cases where $N_m^{\rm u} \leq 10$. Since large $N_m^{\rm u}$ results in large bandwidth, and our goal is to minimize the total bandwidth, $N_m^{\rm u}$ will not be too large. In the proposed search algorithm, we find the optimal solution of problem \eqref{eq:Wtot} in the region $0<N_m^{\rm u}\leq N^{\rm u}_{\max}$, where $N^{\rm u}_{\max}$ is the maximal number of subchannels that can be assigned to each sensor. We will validate that the optimal value of $N_m^{\rm u}$ is not large with numerical results.

Given the value of $N_m^{\rm u}$, according to Property \ref{P:w} and Property \ref{P:fe}, the optimal values of $B_m^{\rm u}$ and $e_m^{\rm u,th}$ that minimize \eqref{eq:su} can be found via binary search method and exact linear search method, respectively. By searching $B_m^{\rm u}$ and $e_m^{\rm u,th}$ with different values of $N^{\rm u}_m \in \{ 1,...,N^{\rm u}_{\max}\}$, the optimal solution of problem \eqref{eq:su} can be obtained, denoted as $\{N^{\rm u*}_m, B^{\rm u*}_m, e^{\rm u,th*}_m \}$. To solve problem \eqref{eq:Wtot}, we need to solve problem \eqref{eq:su} for $M$ sensors. Hence, the complexity of the proposed algorithm is $O(M N^{\rm u}_{\max})$. The details of the algorithm are provided in Table I.

\renewcommand{\algorithmicrequire}{\textbf{Input:}}
\renewcommand{\algorithmicensure}{\textbf{Output:}}
\begin{table}[htb]\small
\vspace{-0.2cm}
\caption{Algorithm to Solve Problem \eqref{eq:su}}
\vspace{-0.8cm}
\begin{tabular}{p{16cm}}
\\\hline
\end{tabular}
\vspace{-0.2cm}
\begin{algorithmic}[1]
\REQUIRE $N^{\rm u}_{\max}$, $T_{\rm f}$, $b$, $N_0$, $N_\mathrm{t}$, $\alpha^{\rm u}_m$, $P^{\rm u}_{\max}$, and accuracy requirement of binary search method $\delta_b$.
\ENSURE $N^{\rm u*}_m$, $B^{\rm u*}_m$, and $e^{\rm u,th*}_m$.
\STATE $N^{\rm u}_m := 1$
\WHILE{$N^{\rm u}_m \leq N^{\rm u}_{\max}$}
\STATE Set $B_{\rm lb} := 0$, $B_{\rm ub} := W_{\rm c}$, $B^{\rm u}_{\rm bs} := 0.5(B_{\rm lb}+B_{\rm ub})$.
\WHILE{$B_{\rm ub} - B_{\rm lb} > \delta_b$}
\STATE Apply exact linear search method to find $e^{\rm u}_{\rm bs}$ that minimizes $f_m^{\rm u}(N^{\rm u}_m,B^{\rm u}_{\rm bs}, e^{\rm u}_{\rm bs})$.
\IF{$f_m^{\rm u}(N^{\rm u}_m,B^{\rm u}_{\rm bs}, e^{\rm u}_{\rm bs}) > \varepsilon^{\rm u}$}
\STATE $B_{\rm lb} := B_{\rm bs}$, $B_{\rm bs} := 0.5(B_{\rm lb}+B_{\rm ub})$.
\ELSE
\STATE $B_{\rm ub} := B_{\rm bs}$, $B_{\rm bs} := 0.5(B_{\rm lb}+B_{\rm ub})$.
\ENDIF
\ENDWHILE
\IF{$f_m^{\rm u}(N^{\rm u}_m,B^{\rm u}_{\rm bs}, e^{\rm u}_{\rm bs}) \leq \varepsilon^{\rm u}$}
\STATE $B^{\rm u}_m(N^{\rm u}_m) := B^{\rm u}_{\rm bs}$ and $e^{\rm u,th}_m(N^{\rm u}_m):=e^{\rm u}_{\rm bs}$.
\ELSE
\STATE $B^{\rm u}_m(N^{\rm u}_m) := \text{NaN}$ and $e^{\rm u,th}_m(N^{\rm u}_m):= \text{NaN}$.
\ENDIF
\STATE $N^{\rm u}_m := N^{\rm u}_m+1$.
\ENDWHILE
\STATE $N_m^{\rm u*} := \mathop {\arg }\limits_{N^{\rm u}_m} \min N^{\rm u}_m\left[{B^{\rm u}_m}\left( {N^{\rm u}_m} \right)/B_0\right]^+B_0$.
\STATE $B^{\rm u *}_m := \left[{B^{\rm u}_m}\left( {N^{\rm u*}_m} \right)/B_0\right]^+B_0$, $e_m^{\rm u,th*} := e^{\rm u,th}_m(N_m^{\rm u *})$.
\RETURN $N_m^{\rm u*}, B^{\rm u*}_m, e_m^{\rm u,th*}$.
\end{algorithmic}
\vspace{-0.4cm}
\begin{tabular}{p{16cm}}
\\
\hline
\end{tabular}
\vspace{-1.0cm}
\end{table}

\subsubsection{DL subchannel assignment}
The optimal DL subchannel assignment that minimizes the required DL bandwidth can be obtained by solving the following problem:
\begin{align}
\mathop {\mathop {\min}\limits_{N^{\rm d},{B^{\rm d}},e^{\rm d,th}_{k_{\min}}} } & \frac{D^{\rm d}}{T_{\rm f}}{E_{\rm B}^+}{N^{\rm d}}{B^{\rm d}} \label{eq:WtotDL}\\
\text{s.t.}\;& 0 < {B^{\rm d}} \le {W_{\rm c}},\label{eq:WcDL}\tag{\theequation a} \\
& 0 < N^{\rm d}, N^{\rm d} \in {\mathbb{Z}}, \label{eq:NhDL}\tag{\theequation b}\\
& \text{and}\;\eqref{eq:DLTR},\nonumber
\end{align}
where $F^{-1}_{\rm R}$ is removed from the objective function since it does not change the optimal solution.
Similar to $f_m^{\rm u}(N^{\rm u}_m,B^{\rm u}_m, e^{\rm u,th}_m)$ in \eqref{eq:ULTr}, we can prove that $f_{k_{\min}}^{\rm d}(N^{\rm d},B^{\rm d}, e^{\rm d,th}_{k_{\min}})$ in \eqref{eq:DLTR} also satisfies Property \ref{P:fw}, Property \ref{P:w} and Property \ref{P:fe}. The proofs are similar to that in Appendices \ref{App:Appendix_P1}, \ref{App:Appendix_P2} and \ref{App:Appendix_P3}, and hence are omitted for conciseness. Therefore, the solution of problem \eqref{eq:WtotDL} can also be found with the algorithm in Table I, and is denoted as $\{N^{\rm d*}_m,B^{\rm d*}_m, e^{\rm d,th*}_m\}$.

\subsection{Delay Components Optimization}
To show how to optimize the delay components and to reduce complexity in optimization, we first analyze the relations between the required bandwidth and the delay components.

\subsubsection{Increasing queueing delay bound} From \eqref{eq:sreq}, we can directly obtain that the required DL service rate decreases with $D^{\rm q}$. As a result, the DL bandwidth decreases with $D^{\rm q}$.

\subsubsection{Increasing UL transmission delay}
In order to show the relationship between UL bandwidth and UL transmission delay, we compare two systems with different UL transmission delays $\hat{D}^{\rm u} < \tilde{D}^{\rm u}$. With $2T_{\rm f}$ delay caused by control signaling, the transmission delay is $\hat{D}^{\rm u}-2T_{\rm f}$ (or $\tilde{D}^{\rm u}-2T_{\rm f}$). Denote $\hat{\bf{1}}_m$ and $\tilde{\bf{1}}_m$ as the indicator functions that indicate whether the $m$th sensor is active in the first and the second systems, respectively. Given the probability that the $m$th sensor requests to transmit a packet in a certain frame as $p_m$, the probabilities that the $m$th sensor is active can be expressed as ${\mathbb{E}}(\hat{\bf{1}}_m)= \frac{\hat{D}^{\rm u}-2T_{\rm f}}{T_{\rm f}}p_m$ and ${\mathbb{E}}(\tilde{\bf{1}}_m) = \frac{\tilde{D}^{\rm u}-2T_{\rm f}}{T_{\rm f}}p_m$, respectively. Then,
\begin{align}
\frac{{T_{\rm f}}{\mathbb{E}}(\hat{\bf{1}}_m)}{\hat{D}^{\rm u}-2T_{\rm f}} = p_m = \frac{{T_{\rm f}}{\mathbb{E}}(\tilde{\bf{1}}_m)}{\tilde{D}^{\rm u}-2T_{\rm f}}. \label{eq:indicator}
\end{align}

The bandwidth for UL transmission in \eqref{eq:Wtot} is hard to analyze since $M_{\rm a}^{\rm th}$ has no closed-form expression. To obtain some useful insights, we study the average bandwidth for UL transmission.
\begin{prop}\label{Pp:UL}
\emph{Increasing $D^{\rm u}$ can reduce the required average bandwidth for UL transmission.}
\begin{proof}
\emph{See proof in Appendix \ref{App:Appendix_P5}.}
\end{proof}
\end{prop}

Although the average bandwidth cannot reflect the bandwidth requirement directly, the simulations in the next section validate that the minimal UL bandwidth decreases with $D^{\rm u}$.

\subsubsection{Increasing DL transmission time} Based on the following proposition for DL transmission, we can obtain a different conclusion from UL transmission.
\begin{prop}\label{Pp:DL}
\emph{If constraint \eqref{eq:WcDL} is inactive (i.e., ${B^{\rm d}} < {W_{\rm c}}$), the required minimal bandwidth for DL transmission does not change with $D^{\rm d}$. }
\begin{proof}
\emph{See proof in Appendix \ref{App:Appendix_P6}.}
\end{proof}
\end{prop}

Given the number of subchannels for each packet transmission, $B^{\rm d}$ increases as $D^{\rm d}$ decreases. Whether constraint \eqref{eq:WcDL} is inactive or not depends on the number of antennas at each BS, the radius of each cell and communication environment. If $N_{\rm t}$ is large or the radius of a cell is small, then $B^{\rm d} < W_{\rm c}$ even when $D^{\rm d} = T_{\rm f}$. We will provide related numerical results in the next section.

\subsubsection{Joint optimization of the three delay components}
The above analysis shows that given E2E delay, the tradeoff between delay components lead to a tradeoff between UL bandwidth and DL bandwidth. To minimize the total bandwidth, we need to optimize the delay components.

For any given $D^{\rm u}$ $D^{\rm d}$ and $D^{\rm q}$, by solving problem \eqref{eq:Wtot} and problem \eqref{eq:WtotDL}, we can obtain the optimal bandwidth assignment policy and the minimized total bandwidth, which are denoted as $\Phi^*(D^{\rm u},D^{\rm d},D^{\rm q}) \triangleq (N^{\rm u*}_m, B^{\rm u*}_m, e^{\rm u,th*}_m,N^{\rm d*},B^{\rm d*}, e^{\rm d,th*}_{k_{\min}})$ and $B_{\rm tot}(\Phi^*(D^{\rm u},D^{\rm d},D^{\rm q}))$, respectively. Since the optimal bandwidth assignment policy depends on the delay components, $\Phi^*(\cdot)$ and $B_{\rm tot}(\cdot)$ are functions of the delay components. To obtain the optimal delay components, we search the values of $D^{\rm u}$, $D^{\rm d}$, and $D^{\rm q}$. Since the possible values of $D^{\rm u}$, $D^{\rm d}$, and $D^{\rm q}$ in problem \eqref{eq:aveBand} are finite, it is not hard to obtain $D^{\rm u*}$, $D^{\rm d*}$, and $D^{\rm q*}$ that minimize $B_{\rm tot}(\Phi^*(D^{\rm u},D^{\rm d},D^{\rm q}))$ with the exhaustive search method.

For large $N_{\rm t}$ or small cells, $B^{\rm d} < W_{\rm c}$. According to Proposition \ref{Pp:DL}, $D^{\rm d*} = T_{\rm f}$. We only need to search $D^{\rm u}$ and $D^{\rm q}$ under the constraint $D^{\rm u}+D^{\rm q} \leq D_{\max}-D^{\rm b} - T_{\rm f}$. Further considering that the bandwidth is minimized when $D^{\rm u}+D^{\rm q} = D_{\max} -D^{\rm b}- T_{\rm f}$, we only need to search $D^{\rm u}$ in $(0,D_{\max} - T_{\rm f})$, which is one-dimensional searching, and hence the complexity is not high.

\subsection{Optimality of the Two-step Method}
To show that $D^{\rm u*}$, $D^{\rm d*}$, $D^{\rm q*}$ and $\Phi^*(D^{\rm u*},D^{\rm d*},D^{\rm q*})$ is the optimal solution of problem \eqref{eq:aveBand}, we need the following proposition.

\begin{prop}\label{Pp:optimal}
\emph{For an arbitrary solution of problem \eqref{eq:aveBand}, $\tilde{D}^{\rm u}$, $\tilde{D}^{\rm d}$, $\tilde{D}^{\rm q}$ and $\tilde{\Phi}(\tilde{D}^{\rm u},\tilde{D}^{\rm d},\tilde{D}^{\rm q})$, we always have $B_{\rm tot}(\Phi^*(D^{\rm u*},D^{\rm d*},D^{\rm q*})) \leq B_{\rm tot}(\tilde{\Phi}(\tilde{D}^{\rm u},\tilde{D}^{\rm d},\tilde{D}^{\rm q})).$}
\begin{proof}
	See proof in Appendix \ref{App:Appendix_Optimal}.
\end{proof}
\end{prop}

This suggests that if both the solution for bandwidth assignment policy and the solution for delay components are global optimal, then the two-step method gives rise to a global optimal solution of problem \eqref{eq:aveBand}. The global optimal bandwidth assignment policy can be obtained by the algorithm in Table I when $W_{\rm c} \leq {B}^{\rm u,min}_m $ and $g^{\rm u,th}_m < N_{\rm t}-1$. Otherwise, we need to use exhaustive searching to find the global optimal solution.

\section{Simulation and Numerical Results}
In this section, we validate the analyses and demonstrate the required bandwidth to support URLLC. With simulation results, we show the impact of minimizing the upper bound of the required total bandwidth and illustrate the performance gain with jointly UL and DL configuration, where arrivals of packet at each sensor are generated by simulation.
 With numerical results, we show the impact of the upper bound in \eqref{eq:UBe} on the bandwidth required by each sensor and illustrate the optimal number of subchannels allocated to each sensor.


\begin{table}[htbp]
	\scriptsize
	\renewcommand{\arraystretch}{1.3}
	\caption{Simulation Parameters \cite{Gerhard2014The,3GPP2010Shadowing,Mehdi2013Performance,3GPP2012MTC}}
	\begin{center}\vspace{-0.4cm}
		\begin{tabular}{|p{5.5cm}|p{1.5cm}||p{5.3cm}|p{1.7cm}|}
			\hline
			Number of BSs & $3$ & Number of sensors
			$M$ & $3000$  \\\hline
			Maximal transmit power of a sensor $P^{\rm u}_{\max}$& $23$~dBm & Maximal transmit power of a BS $P^{\rm d}_{\max}$ & $46$~dBm  \\\hline
			Overall packet loss probability  $\varepsilon _{\max}$ & $10^{-7}$ & Latency in radio access network $D_{\max}-D^{\rm b}$ & $1$~ms  \\\hline
			Frame duration $T_{\rm f}$ & $0.1$ms & Backhaul delay $D^{\rm b}$ & $0.1$~ms  \\\hline
			Single-sided noise spectral density $N_0 $& $-174$~dBm/Hz & Coherence bandwidth $W_{\rm c}$ & $0.5$~MHz  \\\hline
			Maximal number of subchannels allocated to each packet $N^{\rm u}_{\max}$ and $N^{\rm d}_{\max}$& $10$ & Average packet rate generated by each sensor & $100$~packets/s  \\\hline
		\end{tabular}
	\end{center}
	\vspace{-1cm}
\end{table}

Simulation Parameters are listed in Table II. Since broadcast is used in DL transmission, the required bandwidth does not change with the number of users. To ensure the QoS requirement of all users, we only consider the users with the worst large-scale channel gains, i.e., the users located at the edge of each cell. The path loss model is $-10\lg(\alpha_m)=35.3+37.6 \lg(d_m)$, where $d_m$~$\text{(m)}$ is the distance between sensors and the BSs they associated with. $d_m$ is uniformly distributed in $[50,r]$~m, where $r$ is the radius of each cell. We only consider the scenarios where $d_m > 50$~m, because the large-scale channel gain $\alpha^{\rm u}_m$ decreases with $d_m$, and more resources are needed to guarantee QoS requirement with larger sensor-BS distance.

We solve problem \eqref{eq:WtotDL} with different values of $N_{\rm t}$ and cell size to show when the required bandwidth of each subchannel is less than the channel coherence bandwidth. Our results show that if the radius of each cell is $100$~m, then ${B^{\rm d*}} < {W_{\rm c}}$ when $N_{\rm t} \geq 4$, and if the radius is $250$~m, then ${B^{\rm d*}} < {W_{\rm c}}$ when $N_{\rm t} \geq 8$. In the rest of this section, the radius of each cell is set to be $250$~m and $N_{\rm t} \geq 8$. As a result, Proposition \ref{Pp:DL} holds. In other words, required minimal bandwidth for DL transmission dose not change with $D^{\rm d}$. The required total bandwidth is minimized when $D^{\rm d*} = T_{\rm f}$.

\begin{figure}[htbp]
	\vspace{-0.3cm}
	\centering
	\begin{minipage}[t]{0.6\textwidth}
		\includegraphics[width=1\textwidth]{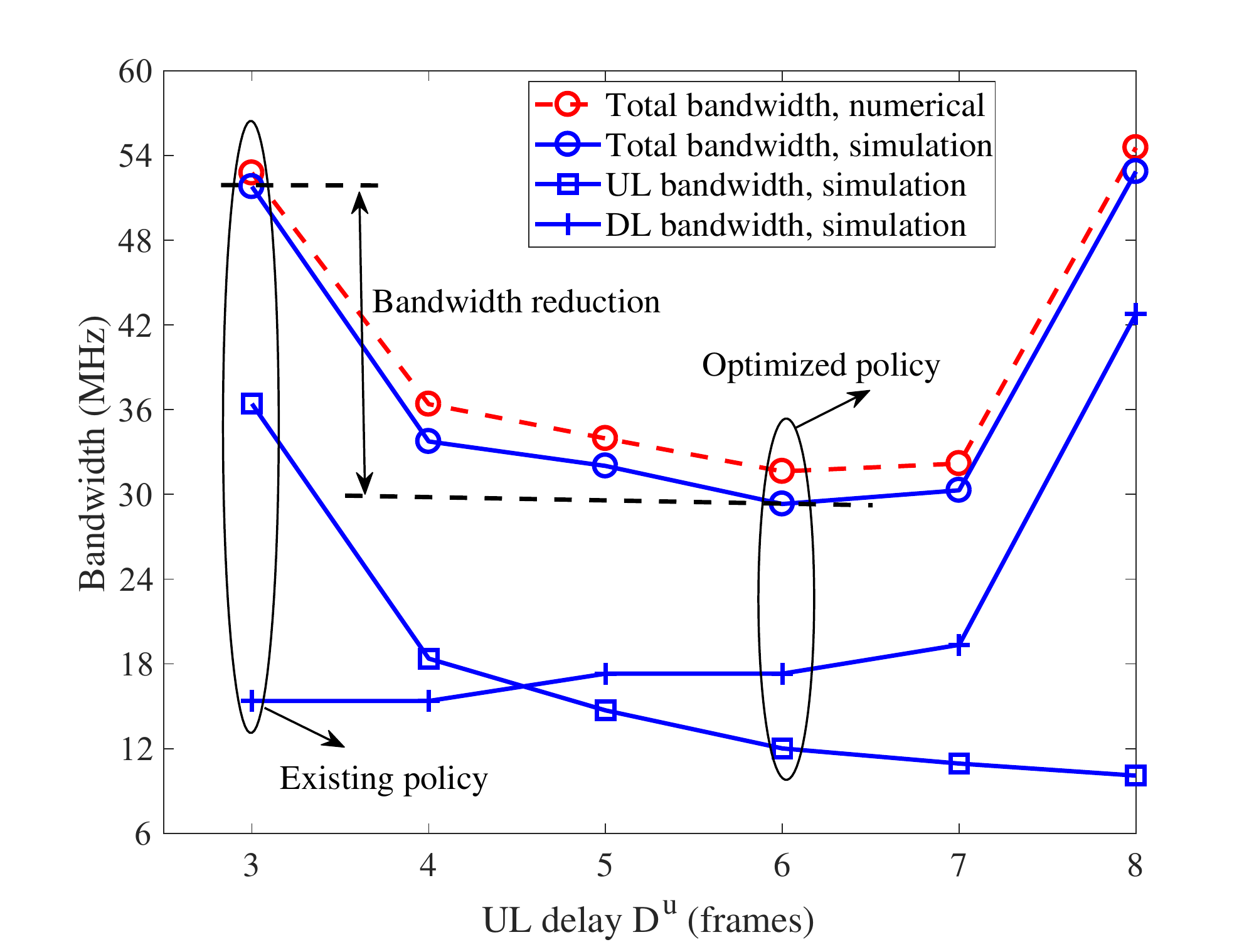}
	\end{minipage}
	\vspace{-0.2cm}
	\caption{Required bandwidth versus delay components, where $\varepsilon^{\rm u}=\varepsilon^{\rm d}=\varepsilon^{\rm q}=\varepsilon_{\max}/3$, and $N_{\rm t} = 8$.}
	\label{fig:Dmax}
	\vspace{-0.5cm}
\end{figure}
Figure \ref{fig:Dmax} shows the required bandwidth with different delay components. The numerical results are obtained by solving problem \eqref{eq:aveBand} with different values of $D^{\rm u}$ and $D^{\rm q}$, where $D^{\rm d*} = T_{\rm f}$ and ${\varepsilon _{\rm M}} = \Pr\{M_{\rm a} > M_{\rm a}^{\rm th}\} = 10^{-15}$ (i.e., ${\varepsilon _{\rm M}} \ll \varepsilon_{\max}$). $M_{\rm a}$ can be approximated by a Poisson process with mean $\lambda(D^{\rm u}-2T_{\rm f})/T_{\rm f}$, where $\lambda$ average number of packets generated by $M$ sensors in one frame. Based on this distribution, we can obtain $M_{\rm a}^{\rm th}$ in \eqref{eq:aveBand} from $\Pr\{M_{\rm a} > M_{\rm a}^{\rm th}\} = 10^{-15}$. To show the impact of minimizing the upper bound of the required total bandwidth, we also provide the simulation results. To obtain the simulation result, we first compute the total bandwidth in \eqref{eq:ob1} achieved by the optimal transmission policy during $10^6$~frames. The maximal total bandwidth in the $10^6$~frames is the required total bandwidth to ensure the QoS requirement and is shown in Figure \ref{fig:Dmax}. We can see that the total bandwidth obtained via numerical results is higher than that obtained via simulation results, and the gap between them is small. This means that the objective function \eqref{eq:aveBand} is a tight upper bound of the total bandwidth in \eqref{eq:ob1}. To show the gain of jointly optimizing the delay components, we provide the results with an existing policy, where the UL data transmission finished in each frame \cite{She2016GCworkshop}. With the existing policy, the required bandwidth is show in Fig. \ref{fig:Dmax} when $D^{\rm u} = 3 T_{\rm f}$ (two frames are occupied by control signaling). We can see that nearly half of the total bandwidth can be saved by optimizing the delay components. The results also indicate that the maximal bandwidth for UL transmission decreases with $D^{\rm u}$, which agrees with Proposition \ref{Pp:UL}.

\begin{table}[btp]
\vspace{-0.2cm}
\scriptsize
\renewcommand{\arraystretch}{1.3}
\caption{Total bandwidth with different packet loss components}
\begin{center}\vspace{-0.2cm}
\begin{tabular}{|p{3.5cm}|p{1.6cm}|p{1.6cm}|p{1.6cm}|}
\hline
   & $N_{\rm t} = 8$ & $N_{\rm t} = 16$ & $N_{\rm t} = 32$  \\\hline
  Optimal $\varepsilon^{\rm u}$, $\varepsilon^{\rm d}$ and $\varepsilon^{\rm q}$ & $28.6$~MHz & $19.9$~MHz& $16.8$~MHz   \\\hline
  $\varepsilon^{\rm u}=\varepsilon^{\rm d}=\varepsilon^{\rm q}=\varepsilon_{\max}/3$ & $29.3$~MHz & $20.2$~MHz& $17.0$~MHz  \\\hline
\end{tabular}
\end{center}
\vspace{-1.2cm}
\end{table}

Simulation results in Table III show the impact of packet loss probabilities on the required bandwidth. The values of $D^{\rm u}$, $D^{\rm d}$ and $D^{\rm q}$ are set as the optimal values that minimize the total bandwidth in Fig. \ref{fig:Dmax}. The optimal values of $\varepsilon^{\rm u}$, $\varepsilon^{\rm d}$ and $\varepsilon^{\rm q}$ are obtained by exhaustive search in the region $[0,\varepsilon_{\max}]$. To reduce complexity, the accuracy is set to be $0.05\varepsilon_{\max}$. For any given values of $\varepsilon^{\rm u}$, $\varepsilon^{\rm d}$ and $\varepsilon^{\rm q}$, the bandwidth assignment is obtained by solving problem \eqref{eq:su} and problem \eqref{eq:WtotDL}. With the bandwidth assignment, the total bandwidth is obtained via simulation, i.e., the maximal total bandwidth in $10^6$~frames. The results show that the total bandwidth with optimal values of $\varepsilon^{\rm u}$, $\varepsilon^{\rm d}$ and $\varepsilon^{\rm q}$ is very close to that with $\varepsilon^{\rm u}=\varepsilon^{\rm d}=\varepsilon^{\rm q}=\varepsilon_{\max}/3$. This validates Remark \ref{R:epson}.

\begin{figure}[htbp]
	\vspace{-0.3cm}
	\centering
	\begin{minipage}[t]{0.55\textwidth}
		\includegraphics[width=1\textwidth]{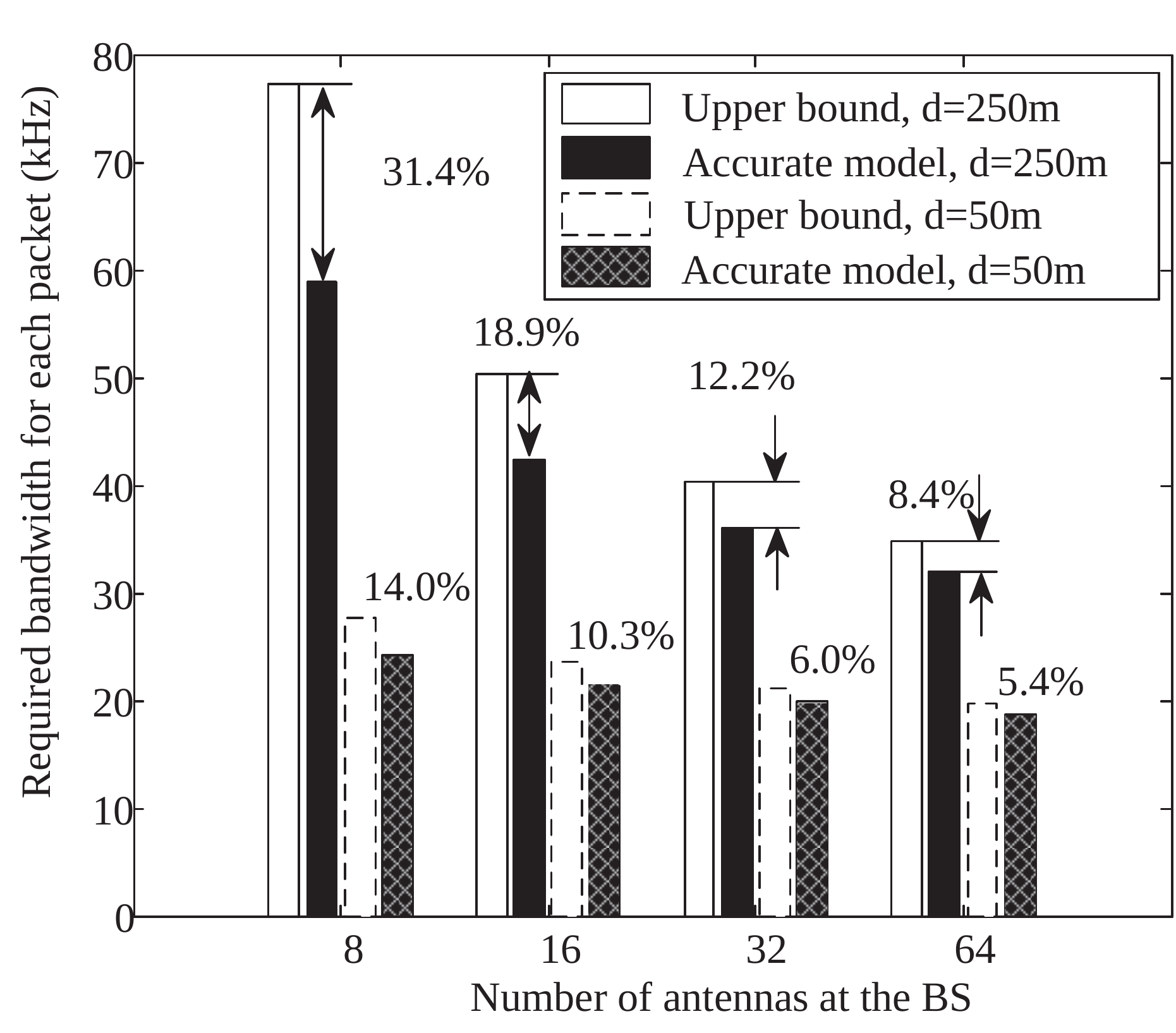}
	\end{minipage}
	\vspace{-0.2cm}
	\caption{Impact of the upper bound in \eqref{eq:UBe} on the required bandwidth, where $\varepsilon^{\rm u} = \varepsilon_{\max}/3$, $D^{\rm u}=6T_{\rm f}$, and $N^{\rm u}_m = 1$.}
	\label{fig:twostate}
	\vspace{-1.0cm}
\end{figure}
The upper bound of packet loss probability in \eqref{eq:UBe} is used to formulate problem \eqref{eq:aveBand}, and hence the bandwidth allocation is conservative. The numerical results in Fig. \ref{fig:twostate} show the impact of the upper bound on the required bandwidth, where UL transmission is considered. For DL transmission, the results are similar. Given the required decoding error probability, the minimal bandwidth with accurate model is obtained by exhaustive searching under constraint $\int_0^\infty  {e_{m,i}^{\rm{u}}{f_g}\left( x \right)} dx \leq \varepsilon^{\rm u}$, where $e_{m,i}^{\rm{u}}$ is given in \eqref{eq:TrE}. The results show that the gap between the minimal bandwidth obtained via the upper bound and that with the accurate model decreases with the number of antennas and increases with the sensor-BS distance. This means that the upper bound has little impact on the resource allocation for macro BSs with a large number of antennas and small BSs with short sensor-BS distance. However, when the number of antennas is small and the sensor-BS distance is large, the upper bound leads to conservative resource allocation. This is because when $g^{\rm u}_{m,i} < g_m^{\rm{u,th}}$, the decoding error probability is much smaller than $1$, setting $e_{m,i}^{\rm{u}} = 1$  leads to conservative resource allocation.

\begin{figure}[htbp]
	\vspace{-0.3cm}
	\centering
	\begin{minipage}[t]{0.6\textwidth}
		\includegraphics[width=1\textwidth]{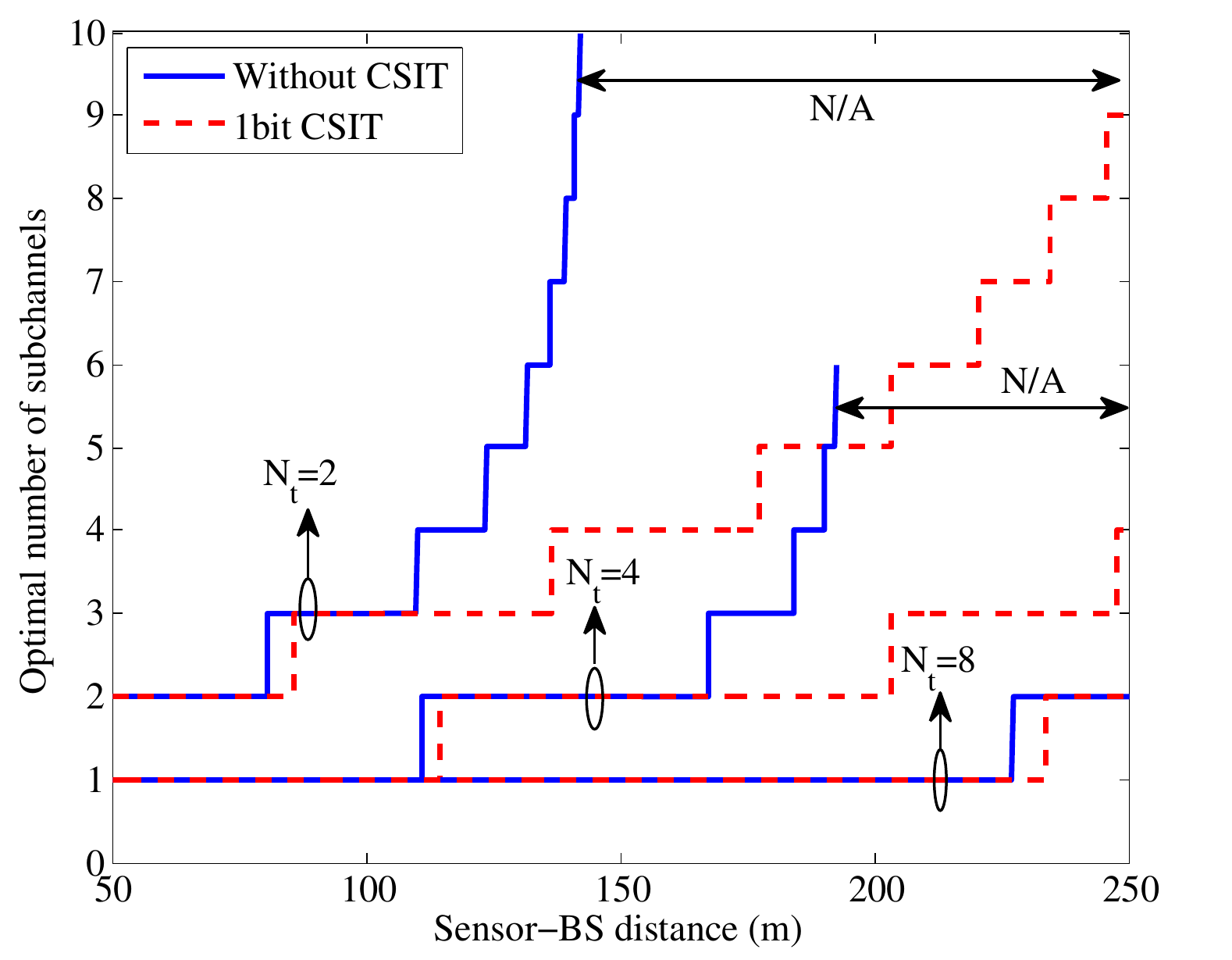}
	\end{minipage}
	\vspace{-0.2cm}
	\caption{Optimal number of links for frequency diversity.}
	\label{fig:Ns}
	\vspace{-0.4cm}
\end{figure}
The optimal number of subchannels assigned to each sensor for UL transmission is illustrated by the numerical results in Fig. \ref{fig:Ns}. We compare two kinds of policies. The first policy does not exploit CSIT, and equally allocates transmit power on the subchannels. With the second policy considered in  \cite{She2016GCworkshop}, one bit information of channel gain on each subchannel is available at the transmitter for resource allocation. With the one bit information, the transmitter knows whether the channel gain on a subchannel is above a threshold that a packet can be decoded successfully with the required probability if the maximal transmit power is allocated to the subchannel. Then, the packet is transmitted on one of the subchannels with good channel. When $N_{\rm t} \geq 16$, $N^*_m = 1$ for all the sensors with any of these two policies (which are not shown in the figure). The results show that CSIT is not helpful for saving bandwidth when the sensor-BS distance is short or the number of active antennas is large.

Availability is another key performance matric for the systems supporting URLLC except the reliability and latency, which is the probability that a system can provide the required QoS  (i.e., $D_{\max}$ and $\varepsilon _{\max}$) for all users \cite{Popovski2014METIS}. Considering that the availability highly depends on shadowing, in the following we provide simulation with shadowing, which follows a lognormal distribution with zero mean and $8$~dB standard deviation \cite{3GPP2010Shadowing}. Since the transmit power of each sensor is limited, we provide the results for UL transmission in Table IV. To obtain the simulation results, we generate the locations and shadowing of $3000$ devices randomly with $10^4$ times. The results show that to guarantee the availability of  $1-10^{-5}$ with a single wireless link, $N_{\rm t}=128$  and $D^{\rm u} = 0.6$~ms. To further improve availability, macro-diversity is an option \cite{Changyang2017TVC}.

\begin{table}[btp]
\vspace{-0.2cm}
\small
\renewcommand{\arraystretch}{1.3}
\caption{Availability when Shadowing is Considered}
\begin{center}\vspace{-0.2cm}
\begin{tabular}{|p{2cm}|p{2cm}|p{2cm}|p{2cm}|p{2cm}|}
\hline
  $N_{\rm t}$ & $16$ & $32$ & $64$ & 128 \\\hline
  $D^{\rm u} = 3 T_{\rm f}$ & $1-5.9\times10^{-2}$ & $1-1.5\times10^{-2}$& $1-3.9\times10^{-3}$ & $1-8.9\times10^{-4}$  \\\hline
  $D^{\rm u} = 4 T_{\rm f}$ & $1-1.2\times10^{-2}$ & $1-1.9\times10^{-3}$& $1-3.1\times10^{-4}$ & $1-5.1\times10^{-5}$  \\\hline
  $D^{\rm u} = 5 T_{\rm f}$ & $1-5.1\times10^{-3}$ & $1-6.7\times10^{-4}$& $1-9.0\times10^{-5}$ & $1-1.1\times10^{-5}$  \\\hline
  $D^{\rm u} = 6 T_{\rm f}$ & $1-2.9\times10^{-3}$ & $1-3.5\times10^{-4}$& $1-4.1\times10^{-5}$ & $1-4.4\times10^{-6}$  \\\hline
\end{tabular}
\end{center}
\vspace{-1.2cm}
\end{table}

\section{Conclusion}
In this paper, we studied joint UL and DL resource configuration to minimize the total bandwidth under strict E2E delay and packet loss probability requirements. A packet delivery mechanism was proposed. In UL transmission, bandwidth is only assigned to the active sensors, and broadcast is used in DL transmission. Channel state information is not available at sensors in UL transmission and not available at the BS in DL transmission. To reduce the required packet rate for ensuring queueing delay at the buffers of the BSs, a statistical multiplexing queueing mode was considered. The total bandwidth required by the mechanism to ensure the E2E delay and overall reliability was minimized by jointly optimizing UL and DL transmission delays, queueing delay and bandwidth assignment. A two-step method was proposed to find the optimal solution of the problem. We first optimized the bandwidth assignment with given delay components and packet loss components. Then, the UL and DL transmission delays and queueing delay were optimized given the E2E delay requirement. Analysis showed that there is a tradeoff between UL and DL bandwidth  and it is necessary to optimize the delay components in order to minimize the total bandwidth. Simulation and numerical results validated our analysis and showed that the joint resource configuration can save half of the total bandwidth comparing with an existing policy, where UL and DL transmission delays are not optimized.


\appendices
\section{Proof of Proposition \ref{P:S}}
\label{App:LS}
\renewcommand{\theequation}{A.\arabic{equation}}
\setcounter{equation}{0}
\begin{proof}
For Poisson arrival process with $\lambda$, the required minimal constant service rate is provided in \eqref{eq:sreq}. From \eqref{eq:sreq}, the required minimal constant service rate of Poisson arrival process with average rate $\tilde{\lambda} = \lambda/L$ is
\begin{align}
\tilde{E}_{\rm B} = \frac{T_{\rm f}\ln(1/\varepsilon^{\rm q})}{D^{\rm q} \ln \left[L \frac{T_{\rm{f}}\ln(1/\varepsilon^{\rm q})}{\lambda D^{\rm q}}+1\right]}\;\text{packets/frame}.\label{eq:sLreq}
\end{align}

To prove $L \tilde{E}_{\rm B} > E_{\rm B}$ ($L = 2,3,...,K$), we only need to prove
\begin{align}
L\ln \left[ \frac{T_{\rm{f}}\ln(1/\varepsilon^{\rm q})}{\lambda D^{\rm q}}+1\right] > \ln \left[L \frac{T_{\rm{f}}\ln(1/\varepsilon^{\rm q})}{\lambda D^{\rm q}}+1\right],\label{eq:L}
\end{align}
which can be obtained by substituting \eqref{eq:sreq} and \eqref{eq:sLreq} into $L \tilde{E}_{\rm B} > E_{\rm B}$. Denote $x = \frac{T_{\rm{f}}\ln(1/\varepsilon^{\rm q})}{\lambda D^{\rm q}}$. Then, \eqref{eq:L} can be equivalently rewritten as
$f_L(x) \triangleq L \ln(x+1) - \ln(Lx+1) > 0, \forall x > 0$. It is easy to show that $f_L(0) = 0$, and $f'_L(x) = \frac{L}{x+1} - \frac{L}{Lx+1}$, which is positive for $L > 1$. Therefore, $f_L(x) > 0, \forall x > 0$. This completes the proof.
\end{proof}

\section{Proof of Property \ref{P:fw}}
\label{App:Appendix_P1}
\renewcommand{\theequation}{B.\arabic{equation}}
\setcounter{equation}{0}
\begin{proof}
By substituting $f_{\rm g}\left( x \right) = \frac{1}{{\left( {{N_\mathrm{t}} - 1} \right)!}}{x^{{N_\mathrm{t}} - 1}}{e^{ - x}}$ into \eqref{eq:ULTr}, we have
\begin{align}
{f_m^{\rm{u}}}(N^{\rm{u}}_m,{B^{\rm{u}}_m},{e^{\rm{u,th}}_m}) = {\left[ {\int_0^{g_m^{{\rm{u,th}}}} {\frac{1}{{\left( {{N_{\rm{t}}} - 1} \right)!}}{x^{{N_{\rm{t}}} - 1}}{e^{ - x}}dx} } + {e^{\rm{u,th}}_m} \right]^{N^{\rm{u}}_m}} .\label{eq:feA}
\end{align}
Denote ${f_{\rm{e}}} = {\int_0^{g_m^{{\rm{u,th}}}} {\frac{1}{{\left( {{N_{\rm{t}}} - 1} \right)!}}{x^{{N_{\rm{t}}} - 1}}{e^{ - x}}dx} }$. To prove property \ref{P:fw}, we only need to prove that ${f_{\rm{e}}}$ first strictly decreases with ${B _m}$ and then increases with ${B _m}$. To this end, we first prove that $f_{\rm{e}}$ strictly increases with $g_m^{\rm u,th}$ and then prove that $g_m^{\rm u,th}$ first strictly decreases with $B^{\rm{u}}_m$ and then increases with ${B _m}$.

From ${f_{\rm{e}}} $, we can obtain that
$\frac{{\partial {f_{{\rm{e}}}}}}{{\partial g_m^{{\rm{u,th}}}}} = \frac{{{{\left( {g_m^{{\rm{u,th}}}} \right)}^{{N_{\rm{t}}} - 1}}{e^{ - g_m^{{\rm{u,th}}}}}}}{{\left( {{N_{\rm{t}}} - 1} \right)!}} > 0$.
As a result, $f_{\rm{e}}$ strictly increases with $g_m^{\rm u,th}$. For notation simplicity, \eqref{eq:thresh} can be rewritten as follows,
\begin{align}
g_m^{\rm u,th} = {C_1{B^{\rm{u}}_m}}\left[\exp\left(\frac{C_2}{B^{\rm{u}}_m}+ \frac{C_3}{\sqrt{{B^{\rm{u}}_m}}}\right)-1\right],\label{eq:gxA}
\end{align}
where $C_1 = \frac{{\phi}N_0}{\alpha^{\rm{u}}_m P^{\rm{u}}_{\max}} > 0$, $C_2 = \frac{b \ln 2}{D^{\rm u} - 2T_{\rm f}}>0$, and $C_3 = \sqrt {\frac{1}{{D^{\rm u} - 2T_{\rm f}}}} f_{\rm Q}^{ - 1}\left( {e^{\rm{u,th}}_m } \right) > 0$.

It is not hard to see that \eqref{eq:gxA} is the same as (19) in \cite{Chengjian2017GC}. According to the proof in Appendix B in \cite{Chengjian2017GC}, $g_m^{\rm u,th}$ first strictly decreases with $B^{\rm{u}}_m$ and then strictly increases with $B^{\rm{u}}_m$, and there is a unique solution of $B^{\rm{u}}_m$ that minimizes $g_m^{\rm u,th}$. This completes the proof.
\end{proof}

\section{Proof of Property \ref{P:w}}
\label{App:Appendix_P2}
\renewcommand{\theequation}{C.\arabic{equation}}
\setcounter{equation}{0}
\begin{proof}
To prove that $f_m^{\rm u}(N^{\rm u}_m,B^{\rm u}_m, e^{\rm u,th*}_m)$ decreases with $B^{\rm u}_m$ in the region  $B^{\rm u}_m \in [0,W_{\rm c}]$, we show that for any $W^{\rm u}_m < \tilde{W}^{\rm u}_m\leq W_{\rm c}$, $f_m^{\rm u}(N^{\rm u}_m,W^{\rm u}_m, e^{\rm u,th*}_m) > f_m^{\rm u}(N^{\rm u}_m,\tilde{W}^{\rm u}_m, \tilde{e}^{\rm u,th*}_m)$, where $\tilde{e}^{\rm u,th*}_m$ is the optimal value of $e^{\rm u,th}_m$ that minimizes $f_m^{\rm u}(N^{\rm u}_m,\tilde{W}^{\rm u}_m, e^{\rm u,th}_m)$ with given $N^{\rm u}_m$ and $\tilde{W}^{\rm u}_m$. According to Property \ref{P:fw}, given $N^{\rm u}_m$ and $e^{\rm u,th*}_m$, we have
\begin{align}
f_m^{\rm u}(N^{\rm u}_m,W^{\rm u}_m, e^{\rm u,th*}_m) > f_m^{\rm u}(N^{\rm u}_m,\tilde{W}^{\rm u}_m, e^{\rm u,th*}_m).\label{eq:feB1}
\end{align}
Since $\tilde{e}^{\rm u,th*}_m$ is the optimal value of $e^{\rm u,th}_m$ that minimizes $f_m^{\rm u}(N^{\rm u}_m,\tilde{W}^{\rm u}_m, e^{\rm u,th}_m)$, we have
\begin{align}
f_m^{\rm u}(N^{\rm u}_m,\tilde{W}^{\rm u}_m, e^{\rm u,th*}_m) \geq f_m^{\rm u}(N^{\rm u}_m,\tilde{W}^{\rm u}_m, \tilde{e}^{\rm u,th*}_m).\label{eq:feB2}
\end{align}
From \eqref{eq:feB1} and \eqref{eq:feB2}, we can obtain that
$f_m^{\rm u}(N^{\rm u}_m,W^{\rm u}_m, e^{\rm u,th*}_m) > f_m^{\rm u}(N^{\rm u}_m,\tilde{W}^{\rm u}_m, \tilde{e}^{\rm u,th*}_m)$. The proof follows.
\end{proof}

\section{Proof of Property \ref{P:fe}}
\label{App:Appendix_P3}
\renewcommand{\theequation}{D.\arabic{equation}}
\setcounter{equation}{0}
\begin{proof}
According to \eqref{eq:feA}, to study the convexity of ${f_m^{\rm{u}}}(N^{\rm{u}}_m,{B^{\rm{u}}_m},{e^{\rm{u,th}}_m})$, we only need to study the convexity of ${f_{\rm{e}}}$. To this end, we first prove that $g_m^{\rm u,th}$ in \eqref{eq:thresh} is convex in ${e^{\rm{u,th}}_m}$. Then, we show that $f_{\rm{e}}$ is an increasing and convex function of $g_m^{\rm u,th}$ when $g_m^{\rm u,th} < N_{\rm t} - 1$.

For the Q-function  ${f_{\rm Q}}\left( x \right) = \frac{1}{{\sqrt {2\pi } }}\int_x^\infty  {\exp \left( { - \frac{{{\tau ^2}}}{2}} \right)} d\tau$, we have ${f'_{\rm Q}}\left( x \right) \buildrel \Delta \over =  - \frac{1}{{\sqrt {2\pi } }}{e^{ - {x^2}/2}} < 0$, and ${f''_{\rm Q}}\left( x \right) = \frac{x}{{\sqrt {2\pi } }}{e^{ - {x^2}/2}} > 0$ when $x > 0$. Thus, when $x > 0$, ${f_{\rm Q}}\left( x \right)$ is a decreasing and convex function. Moreover, ${f_{\rm Q}}\left( x \right) < 0.5, \forall x > 0$. Because $e^{\rm{u,th}}_m <\varepsilon_{\max}<0.5$ that is
true for URLLC applications, and the inverse function of a decreasing and convex function is also convex \cite{boyd}, $f_{\rm Q}^{ - 1}\left( {e^{\rm{u,th}}_m} \right)$ is convex in $e^{\rm{u,th}}_m$, $\forall e^{\rm{u,th}}_m < 0.5$. Denote $z = f_{\rm Q}^{ - 1}\left( {e^{\rm{u,th}}_m} \right)$. Then, $g_m^{\rm u,th}$ in \eqref{eq:thresh} can be rewritten as $
g_m^{{\rm{u,th}}} = {C_4}\left[ {\exp \left( {{C_5} + {C_6}z} \right) - 1} \right]$, where $C_4 = \frac{{{{\phi}N_0}{B^{\rm{u}}_m}}}{{{\alpha^{\rm{u}}_m}{P^{\rm{u}}_{\max}}}}>0$, $C_5 = \frac{{b\ln 2}}{{({D^{\rm u} - 2T_{\rm f}}) {B^{\rm{u}}_m}}}>0$ and $C_6 = \sqrt {\frac{1}{{({D^{\rm u} - 2T_{\rm f}}) {B^{\rm{u}}_m}}}}>0$. It is easy to see that $g_m^{{\rm{u,th}}}$ is an increasing and convex function of $z$. According to the composition rules, $g_m^{\rm u,th}$ is convex in $e^{\rm{u,th}}_m$ \cite{boyd}.

It is not hard to derive that $\frac{{{\partial ^2}{{{f_{{\rm{e}}}}} }}}{{\partial {{\left( {g_m^{{\rm{u,th}}}} \right)}^2}}} = \frac{{{{\left( {g_m^{{\rm{u,th}}}} \right)}^{{N_{\rm{t}}} - 2}}{e^{ - g_m^{{\rm{u,th}}}}}}}{{\left( {{N_{\rm{t}}} - 1} \right)!}}\left( { {N_{\rm t}}} - 1 - g_m^{{\rm{u,th}}} \right).$
When $N_{\rm t} - 1 \geq g_m^{{\rm{u,th}}}$, ${f_{\rm{e}}}$ is increasing and convex in $g_m^{{\rm{u,th}}}$. According to the composition rules, ${f_{\rm{e}}}$ is convex in $e^{\rm{u,th}}_m$, when $N_{\rm t} - 1 \geq g_m^{{\rm{u,th}}}$. This completes the proof.
\end{proof}
\section{Proof of Proposition \ref{Pp:UL}}
\label{App:Appendix_P5}
\renewcommand{\theequation}{E.\arabic{equation}}
\setcounter{equation}{0}
\begin{proof}
Denote the bandwidth for the $m$th sensor in two systems as $\hat{B}_m^{\rm u}$ and $\tilde{B}_m^{\rm u}$, respectively. To keep the decoding error probability identical, the values of $N^{\rm u}_m$, $g^{\rm u,th}_m$, $e^{\rm u,th}_m$ in \eqref{eq:ULTr} are fixed in the two systems. According to \eqref{eq:thresh}, the relationship between $\hat{B}_m^{\rm u}$ and $\tilde{B}_m^{\rm u}$ can be obtained from
\begin{align}
&\hat{B}^{\rm u}_m\left\{ {\exp \left[ {\frac{{b\ln 2}}{{(\hat{D}^{\rm u}-2T_{\rm f}) {\hat{B}^{\rm u}_m}}} + \sqrt {\frac{1}{{(\hat{D}^{\rm u}-2T_{\rm f}) {\hat{B}^{\rm u}_m}}}} f_{\rm Q}^{ - 1}\left( {e^{\rm u,th}_{m} } \right)} \right] - 1} \right\} \nonumber\\
=&\tilde{B}^{\rm u}_m\left\{ {\exp \left[ {\frac{{b\ln 2}}{{(\tilde{D}^{\rm u}-2T_{\rm f}) {\tilde{B}^{\rm u}_m}}} + \sqrt {\frac{1}{{ (\tilde{D}^{\rm u}-2T_{\rm f}) {\tilde{B}^{\rm u}_m}}}} f_{\rm Q}^{ - 1}\left( {e^{\rm u,th}_{m} } \right)} \right] - 1} \right\}.\label{eq:Brelation}
\end{align}
Since in typical scenarios ${B}_m^{\rm u,min} \geq W_{\rm c}$, $g^{\rm u,th}_m$ in \eqref{eq:thresh} strictly decreases with ${B}_m^{\rm u}$ in the region $[0,W_{\rm c}]$. Hence, both left and right hand sides of \eqref{eq:Brelation} decrease with ${B}_m^{\rm u}$. Moreover, by substituting $\tilde{B}^{\rm u}_m = \frac{\hat{D}^{\rm u}-2T_{\rm f}}{\tilde{D}^{\rm u}-2T_{\rm f}}\hat{B}^{\rm u}_m $ into \eqref{eq:Brelation}, the left hand side of \eqref{eq:Brelation} is larger than the right hand side of it. Therefore, to satisfy \eqref{eq:Brelation}, $\tilde{B}^{\rm u}_m < \frac{\hat{D}^{\rm u}-2T_{\rm f}}{\tilde{D}^{\rm u}-2T_{\rm f}}\hat{B}^{\rm u}_m$. From $\tilde{B}^{\rm u}_m < \frac{\hat{D}^{\rm u}-2T_{\rm f}}{\tilde{D}^{\rm u}-2T_{\rm f}}\hat{B}^{\rm u}_m$ and \eqref{eq:indicator}, we have
\begin{align}
{\mathbb{E}}(\sum\limits_{m = 1}^M \tilde{\bf{1}}_m {N}_m^{\rm u} \tilde{B}_m^{\rm u}) < \sum\limits_{m = 1}^M \frac{\tilde{D}^{\rm u}-2T_{\rm f}}{\hat{D}^{\rm u}-2T_{\rm f}}{\mathbb E}(\hat{\bf{1}}_m) {N}_m^{\rm u} \frac{\hat{D}^{\rm u}-2T_{\rm f}}{\tilde{D}^{\rm u}-2T_{\rm f}}\hat{B}_m^{\rm u} = {\mathbb{E}}(\sum\limits_{m = 1}^M \hat{\bf{1}}_m {N}_m^{\rm u} \hat{B}_m^{\rm u}).\nonumber
\end{align}
The proof follows.
\end{proof}

\section{Proof of Proposition \ref{Pp:DL}}
\label{App:Appendix_P6}
\renewcommand{\theequation}{F.\arabic{equation}}
\setcounter{equation}{0}
\begin{proof}
To prove Proposition \ref{Pp:DL}, we need to prove that the minimal DL bandwidth obtained by solving problem \eqref{eq:WtotDL}  does not change with $D^{\rm d}$ when constraint \eqref{eq:WcDL} is inactive. We consider two systems with different DL transmission time, i.e., $\hat{D}^{\rm d} \ne \tilde{D}^{\rm d}$. We refer to problem \eqref{eq:WtotDL} with $\hat{D}^{\rm d}$ and $\tilde{D}^{\rm d}$ as Problem A and Problem B, respectively. Denote the optimal solutions of Problem A and Problem B as $\{\hat{N}^{\rm d*}, \hat{B}^{\rm d*},\hat{e}_{k_{\min}}^{\rm d,th*}\}$ and $\{\tilde{N}^{\rm d*}, \tilde{B}^{\rm d*},\tilde{e}_{k_{\min}}^{\rm d,th*}\}$, respectively. Given transmission duration $\hat{D}^{\rm d}$ and $\{\hat{N}^{\rm d*}, \hat{B}^{\rm d*},\hat{e}_{k_{\min}}^{\rm d,th*}\}$, the threshold in \eqref{eq:threshDL} is denoted as $\hat{g}^{\rm d,th}_{k_{\min}}$.

To prove $\frac{\hat{D}^{\rm d}}{T_{\rm f}}{E_{\rm B}^+}{\hat{N}^{\rm d*}}{\hat{B}^{\rm d*}} = \frac{\tilde{D}^{\rm d}}{T_{\rm f}}{E_{\rm B}^+}{\tilde{N}^{\rm d*}}{\tilde{B}^{\rm d*}}$, we assume they are not equal, and find contradiction. Without loss of generality, we assume $\frac{\hat{D}^{\rm d}}{T_{\rm f}}{E_{\rm B}^+}{\hat{N}^{\rm d*}}{\hat{B}^{\rm d*}} < \frac{\tilde{D}^{\rm d}}{T_{\rm f}}{E_{\rm B}^+}{\tilde{N}^{\rm d*}}{\tilde{B}^{\rm d*}}$.
To this end, we first validate that $\{\hat{N}^{\rm d*}, \frac{\hat{D}^{\rm d}}{\tilde{D}^{\rm d}}\hat{B}^{\rm d*},\hat{e}_{k_{\min}}^{\rm d,th*}\}$ is a feasible solution of problem B. Since $\hat{N}^{\rm d*}$ is a solution of problem A, constraint \eqref{eq:NhDL} is satisfied. Then, we only need to validate that \eqref{eq:DLTR} is satisfied. $\hat{N}^{\rm d*}$ and $\hat{e}_{k_{\min}}^{\rm d,th*}$ are the same in $\{\hat{N}^{\rm d*},\hat{B}^{\rm d*},\hat{e}_{k_{\min}}^{\rm d,th*}\}$ and $\{\hat{N}^{\rm d*}, \frac{\hat{D}^{\rm d}}{\tilde{D}^{\rm d}}\hat{B}^{\rm d*},\hat{e}_{k_{\min}}^{\rm d,th*}\}$. Since $\{\hat{N}^{\rm d*},\hat{B}^{\rm d*},\hat{e}_{k_{\min}}^{\rm d,th*}\}$ is a solution of problem A, constraint \eqref{eq:DLTR} is satisfied with $\hat{N}^{\rm d*}$, $\hat{e}_{k_{\min}}^{\rm d,th*}$ and $\hat{g}^{\rm d,th}_{k_{\min}}$. If $g^{\rm d,th}_{k_{\min}}$ with transmission duration $\tilde{D}^{\rm d}$ and solution $\{\hat{N}^{\rm d*}, \frac{\hat{D}^{\rm d}}{\tilde{D}^{\rm d}}\hat{B}^{\rm d*},\hat{e}_{k_{\min}}^{\rm d,th*}\}$ is the same as $\hat{g}^{\rm d,th}_{k_{\min}}$, then constraint \eqref{eq:DLTR} is satisfied. Substituting $\frac{\hat{D}^{\rm d}}{\tilde{D}^{\rm d}}\hat{B}^{\rm d*}$ and $\hat{e}_{k_{\min}}^{\rm d,th*}$ into  \eqref{eq:threshDL}, we have
\begin{align}
g^{\rm d,th}_{k_{\min}} = &\frac{{\tilde{D}^{\rm d}{\frac{\hat{D}^{\rm d}}{\tilde{D}^{\rm d}}\hat{B}^{\rm d*}}}}{T_{\rm f}}\left\{ {\exp \left[ {\frac{{b\ln 2}}{{\tilde{D}^{\rm d}{\frac{\hat{D}^{\rm d}}{\tilde{D}^{\rm d}}\hat{B}^{\rm d*}}}} + \sqrt {\frac{1}{{\tilde{D}^{\rm d}{\frac{\hat{D}^{\rm d}}{\tilde{D}^{\rm d}}\hat{B}^{\rm d*}}}}} f_{\rm Q}^{ - 1}\left( \hat{e}_{k_{\min}}^{\rm d,th*} \right)} \right] - 1} \right\}\nonumber\\
=&\frac{{\hat{D}^{\rm d}{\hat{B}^{\rm d}}}}{T_{\rm f}}\left\{ {\exp \left[ {\frac{{b\ln 2}}{{\hat{D}^{\rm d} {\hat{B}^{\rm d}}}} + \sqrt {\frac{1}{{\hat{D}^{\rm d} {\hat{B}^{\rm d}}}}} f_{\rm Q}^{ - 1}\left( \hat{e}_{k_{\min}}^{\rm d,th*} \right)} \right] - 1} \right\}=\hat{g}^{\rm d,th}_{k_{\min}}.\label{eq:threshDL2}
\end{align}
Therefore, $\{\hat{N}^{\rm d*}, \frac{\hat{D}^{\rm d}}{\tilde{D}^{\rm d}}\hat{B}^{\rm d*},\hat{e}_{k_{\min}}^{\rm d,th*}\}$ is a feasible solution of problem B.

Given the transmission duration $\tilde{D}^{\rm d}$,
the number of packets that are transmitted simultaneously is $\frac{\tilde{D}^{\rm d}}{T_{\rm f}}{E_{\rm B}^+}$. Therefore, the required bandwidth for DL transmission with $\{\hat{N}^{\rm d*}, \frac{\hat{D}^{\rm d}}{\tilde{D}^{\rm d}}\hat{B}^{\rm d*},\hat{e}_{k_{\min}}^{\rm d,th*}\}$ satisfies $\frac{\tilde{D}^{\rm d}}{T_{\rm f}}{E_{\rm B}^+} \hat{N}^{\rm d*} \frac{\hat{D}^{\rm d}}{\tilde{D}^{\rm d}}\hat{B}^{\rm d*} = \frac{\hat{D}^{\rm d}}{T_{\rm f}}{E_{\rm B}^+}\hat{N}^{\rm d*}\hat{B}^{\rm d*} < \frac{\tilde{D}^{\rm d}}{T_{\rm f}}{E_{\rm B}^+}{\tilde{N}^{\rm d*}}{\tilde{B}^{\rm d*}}$, which contradicts with the assumption that $\{\tilde{N}^{\rm d*}, \tilde{B}^{\rm d*},\tilde{e}_{k_{\min}}^{\rm d,th*}\}$ is the optimal solution of problem B. The proof follows.
\end{proof}

\section{Proof of Proposition \ref{Pp:optimal}}
\label{App:Appendix_Optimal}
\renewcommand{\theequation}{G.\arabic{equation}}
\setcounter{equation}{0}
\begin{proof}
Given the delay components $\tilde{D}^{\rm u}$, $\tilde{D}^{\rm d}$, $\tilde{D}^{\rm q}$, from the first step of the two-step method the optimal subchannel assignment policy is ${\Phi}^*(\tilde{D}^{\rm u},\tilde{D}^{\rm d},\tilde{D}^{\rm q})$, i.e.,
\begin{align}\label{eq:firststep}
B_{\rm tot}({\Phi}^*(\tilde{D}^{\rm u},\tilde{D}^{\rm d},\tilde{D}^{\rm q})) \leq B_{\rm tot}(\tilde{\Phi}(\tilde{D}^{\rm u},\tilde{D}^{\rm d},\tilde{D}^{\rm q})).
\end{align}
According to the second step of the two-step method, the optimal delay components that minimizes $B_{\rm tot}({\Phi}^*({D}^{\rm u},{D}^{\rm d},{D}^{\rm q}))$ are $D^{\rm u*}$, $D^{\rm d*}$ and $D^{\rm q*}$, and hence
\begin{align}\label{eq:secondstep}
B_{\rm tot}({\Phi}^*({D}^{\rm u*},{D}^{\rm d*},{D}^{\rm q*}))\leq B_{\rm tot}({\Phi}^*(\tilde{D}^{\rm u},\tilde{D}^{\rm d},\tilde{D}^{\rm q})).
\end{align}
From \eqref{eq:firststep} and \eqref{eq:secondstep}, we have Proposition \ref{Pp:optimal}. This completes the proof.
\end{proof}

\bibliographystyle{IEEEtran}
\bibliography{ref}

\end{document}